\documentclass[12pt]{iopart}
\usepackage{graphicx}
\usepackage{color}

\usepackage{amssymb}
\def\
crpar{\mbox{$\not\!\partial$}}

\def\half{\frac{1}{2}}

\def\XXint#1#2#3{{\setbox0=\hbox{$#1{#2#3}{\int}$}
     \vcenter{\hbox{$#2#3$}}\kern-.5\wd0}}

\newcommand{\la}{\label}
\newcommand{\be}{\begin{equation}}
\newcommand{\ee}{\end{equation}}
\newcommand{\bea}{\begin{eqnarray}}
\newcommand{\eea}{\end{eqnarray}}

\newcommand{\ba}{\begin{align}}
\newcommand{\ea}{\end{align}}
\newcommand{\1}{\frac{1}{2}}

\def\XXint#1#2#3{{\setbox0=\hbox{$#1{#2#3}{\int}$}
     \vcenter{\hbox{$#2#3$}}\kern-.5\wd0}}

\begin{document}

% \today \hfill {\color{red}\textbf{\underline{Preliminary Draft}}}

\title[On Duality, emergence of Calogero family of models in external potentials, Solitons and Field theory
 \ldots]{Emergence of Calogero family of models in external potentials: Duality, Solitons and Hydrodynamics
% \\ \textit{PAPER 4a}
 }

\author{Manas Kulkarni
%\footnote{E-mail:  alexandre.abanov@sunysb.edu}
}
%\affiliation
\address{International centre for theoretical sciences, Tata Institute of Fundamental Research, Bangalore - 560089, India}

\author{Alexios Polychronakos
%\footnote{E-mail:  alexandre.abanov@sunysb.edu}
}
%\affiliation
\address{Physics Department, City College of the CUNY, New York, NY 10031, USA  and }
%\address{}
\address{The Graduate Center, The City University of New York, New York, NY 10016, USA}
%\ead{alexandre.abanov@sunysb.edu}
%}
%\affiliation
%\address{Physics Department, City College of the CUNY, New York, NY 10031, USA  and }
%\address{}
%\address{The Graduate Center, The City University of New York, New York, NY 10016, USA}
%\ead{alexandre.abanov@sunysb.edu}

%
%\address{}
%
%\address{Department of Condensed Matter Physics and Material Science,
%Brookhaven National Laboratory, Upton, NY-11973. }

%\ead{alexandre.abanov@sunysb.edu}

\begin{abstract}
We present a first-order formulation of the Calogero model in external potentials in terms of a generating function, which simplifies
the derivation of its dual form. Solitons naturally appear in this formulation as particles of negative mass. Using this method, we
obtain the dual form of Calogero particles in external quartic, trigonometric and hyperbolic potentials, which were known to be
integrable but had no known dual formulation. We derive the corresponding soliton solutions, generalizing earlier results for
the harmonic Calogero system, and present numerical results that demonstrate the integrable nature of the soliton motion.
We also give the collective fluid mechanical formulation of these models and derive the corresponding fluid
soliton solutions in terms of meromorphic fields, commenting on issues of stability and integrability.

\end{abstract}

\maketitle
%\pagenumbering{Roman}

\tableofcontents

\section{Introduction}
%%%%%%%%%%%%%%%%%%%%%%%%%%%%%%%%%%%%%%%%

The Calogero model and its various generalizations is one of the most studied integrable systems in physics and mathematics
\cite{Calogero-1969,Sutherland-1971,Moser}. In its most basic form it describes $N$ identical non-relativistic particles in one dimension interacting through two-body inverse-square potentials in the presence of an external harmonic potential.
The Hamiltonian of the rational Calogero model in a harmonic trap reads
\bea
    {\cal H} &=& \frac{1}{2}\sum_{j=1}^{N}\big(p_j^{2}+\omega^2 x_j^2 \big)
   +\1 \sum_{j,k=1; j\neq k}^{N} \frac{g^{2}}{(x_{j}-x_{k})^{2}},
     \la{hCM} %\\
%       &=&  \1 \sum_{j=1}^N \left| p_j- i\omega x_j
%       + ig \sum_{k=1; k\neq j}^{N}\frac{1}{x_j-x_k}  \right|^2
%       + \frac{\omega g}{4} N(N-1)
  \la{Hsquare}
\label{Eq1}
\eea
where $x_{j}$ are the coordinates of the particles, $p_{j}$ are their canonical momenta, and $g$ is the coupling constant. We normalized the mass of the particles to be unity. This model appears in many branches of physics and mathematics and has connections and relevance to fractional statistics, fluid mechanics, spin chains, two-dimensional gravity, strings in low dimensions etc; As a result, it has been studied extensively (see \cite{OlshaPer,Perelomov-book, Sutherland-book, APreview} for reviews and a comprehensive list of references).

The above model is classically and quantum mechanically integrable, and can be obtained as a reduction of a matrix system.
Several generalizations are also integrable, involving hyperbolic, trigonometric or elliptic mutual potentials, and also general
external potentials of quartic type or corresponding trigonometric type \cite{i3,APnew1,APnew2}. Moreover, integrable generalizations where the particles carry internal degrees of freedom have been proposed \cite{Kawakami1,Kawakami2,HaHa,MinahanAP,HiWa}.

%{\color{blue}{N. Kawakami, Phys. Rev. B46, 1005 and 3191 (1992).\\
%
%Z.N.C. Ha and F.D.M. Haldane, Phys. Rev. B46, 9359 (1992).\\
%
%J.A. Minahan and A. P. Polychronakos, Phys. Lett. B302, 265 (1993) [arXiv:hep-th/9206046].\\
%
%K. Hikami and M. Wadati, Phys. Lett. A173, 263 (1993).}}

A particular scaling limit (the ``freezing trick") then leads to integrable spin chains with long-range interactions, including the
Haldane-Shastry spin model as well as the non-translation invariant harmonic spin chain \cite{APspinchain, APspinchain1}.
%{\color{blue}{Lattice integrable systems of Haldane-Shastry type
%Alexios P. Polychronakos (Natl. Tech. U., Athens & CERN). Oct 1992. 8 pp.
%Published in Phys.Rev.Lett. 70 (1993) 2329-2331.\\
%Exact spectrum of SU(n) spin chain with inverse square exchange
%lexios P. Polychronakos (CERN). Oct 1993. 15 pp.
%Published in Nucl.Phys. B419 (1994) 553-566}}

A remarkable fact is that the hydrodynamic limit $N\to\infty$ of the system (\ref{Eq1}) can be found exactly using the methods of collective field theory \cite{JevickiSakita,Sakita-book,Jevicki-1992} or using the methods of Ref. \cite{2005-AbanovWiegmann,2009-AbanovBettelheimWiegmann}. This is quite nontrivial, as writing classical microscopic Hamiltonians in collective fluid mechanical
variables usually involves several approximations. By contrast, the Calogero fluid theory reproduces the dynamics of the many-body
system to all orders in $1/N$, with any corrections being of nonperturbative nature.

The integrability and other rich properties of the underlying particle systems suggest that the corresponding fluid mechanical equations are also integrable and point to the existence of soliton solutions. This has, indeed, been verified for the free ($\omega =0$) Calogero model \cite{1995-Polychronakos}
as well as the periodic trigonometric (Sutherland) model \cite{1995-Polychronakos}. A remarkable method of finding multi-soliton solutions
was recently proposed \cite{kul1,kul2}, relying on a ``dual" representation of the model, where the dual particles play the role of
``soliton variables''. The existence of such a dual form is far from obvious, and the demonstration that it reduces to the usual
Calogero model is quite involved.

The main goal of this paper is to present a first-order formalism based on a generating function (``prepotential")
that makes the self-dual version of the model
apparent and greatly simplifies the derivation of its connection to the second-order Calogero equations of motion. Based on this formalism, generalizations are proposed that can, in principle, admit non-identical particles with different masses and particle-dependent
interactions.
Conditions of stability and reality, subsequently, reduce the system to the dual formulation of the usual Calogero model and
its trigonometric and hyperbolic generalizations, but with more general external potentials which, interestingly, turn out to be the integrable quartic or trigonometric potentials found before  \cite{i3,APnew1,APnew2}. In this formulation, soliton solutions can be identified in these more general potentials. Remarkably, solitons behave as regular Calogero particles but with negative mass and complex coordinates.
From this starting point, a similar finite dimensional reduction can also be performed in the hydrodynamic model, with the fluid motion parametrized in terms of a finite number of complex parameters representing soliton positions and speeds. Issues of stability and soliton condensation are also discussed %% {\color{red} [should we discuss soliton condesnation]}.

The paper is organized as follows:
In Section 2, we introduce our first-order formalism and derive the general form of two-body and external potentials that can be described this way. The solution of the functional equations that appear and derivation of the specific potentials are delegated to the appendices. We examine the stability and reality conditions and establish the appearance of the generalized quartic-type potentials and solitons.
In Section 3, we focus on systems with quartic external potentials and derive their dual formulation and soliton solutions. We comment on the mapping of the soliton problem to an electrostatic problem and present numerical solutions that demonstrate the integrable
nature of the soliton solutions.
In Section 4, we deal with the collective field and fluid mechanical formulation of these models and present the corresponding
soliton solutions in terms of meromorphic fields.
Finally, in Section 5 we state our conclusions and point to directions of future investigation.

%\section{General First-Order Formalism of Interacting Systems}

\section{General First-Order Formalism of Interacting Systems}

In this section, we will formulate the first-order dual equations of motion for a dynamical system of particles
in terms of a generating function (prepotential). This formulation greatly simplifies the proof of 
equivalence of the dual equations with the second-order equations of a Calogero-like system and allows for
generalizations involving more general two-body and external potentials.

The starting point is a system of $n$ particles on the line with coordinates $x_a$,
$a = 1, \dots , n$, obeying the {\it first-order} equations of motion
\bea
\label{xdot1}
m_{a}\dot{x}_{a}=\partial_{a}\Phi
\eea
with $\Phi$ a function of the $x_a$, $\partial_a \equiv {\partial \over \partial x_a}$,
and $m_a$ a set of constant ``masses". Taking another time derivative
we obtain
\bea
m_{a}\ddot{x}_{a}&=& \sum_b \partial_b \partial_a \Phi \, \dot{x}_b = \sum_b \partial_{a}\partial_{b}\Phi\frac{1}{m_{b}}\partial_{b}\Phi \nonumber \\&=&\partial_{a}\left[\sum_{b}\frac{1}{2 m_{b}}\left(\partial_{b}\Phi\right)^{2}\right]
\eea
This has the form of a standard equation of motion for particles of mass $m_a$ inside a potential
\be
V = -\sum_{a}\frac{1}{2 m_{a}}\left(\partial_{a}\Phi\right)^{2}
\label{V}
\ee
such that 
\bea
m_{a}\ddot{x}_{a}&=&-\frac{\partial V}{\partial x_a}
\eea
We wish the potential to contain only one-body and two-body terms. Further, the interactions (two-body terms)
should depend only on the relative particle distance. So we choose a $\Phi$ of the general form
\bea
\Phi=\half \sum_{a\neq b}F_{ab}(x_{ab})+ \sum_a W_a(x_a) ~,~~~ x_{ab} := x_a - x_b
\eea
By symmetrizing the sum, the $F_{ab}$ can be chosen to satisfy $F_{ab} (x) = F_{ba} (-x)$. 
This symmetry ensures that the function $F_{ab}$ depends on the difference between $x_a$ and $x_b$ without caring about
the order of particles. This gives us
\bea
\label{derphi}
\partial_{a}\Phi= \sum_{b}f_{ab}(x_{ab})+w_{a}(x_{a})
\eea
where we defined
\bea
f_{ab} (x) = -f_{ba} (-x) = F'_{ab} (x) ~,~~~ w_a (x)= W'_a (x)
\eea
We look for conditions for $f_{ab}$ and $w_a$ such that the potential contains only
one- and two-body terms. We first examine the case of no external potential.

\subsection{The case $W_a (x)=0$}
At the moment, let us ignore $W_a (x_a)$, i.e, consider the case of no external potential. The expression 
for the potential $V$ then is
\bea
V =-\sum_{b\neq c,d}\frac{1}{2 m_{b}}f_{bc}(x_{bc})f_{bd}(x_{bd})
\eea
Let us define from here on the shorthand (and similarly for other functions)
\bea
f_{ab} \equiv f_{ab} (x_{a}-x_{b}) ~,~~{\rm satisfying}~~ f_{ab} = - f_{ba}
\eea
We also define renormalized functions $\tilde{f}_{ab}$,
\bea
f_{ab}=m_{a}m_{b}\tilde{f}_{ab}
\eea
Then the above potential becomes
\bea
V &=&-\frac{1}{2}\sum_{b\neq c,d} m_{b}m_{c}
m_{d}\tilde{f}_{cb}\tilde{f}_{db} \nonumber \\
&=&-\half \sum_{b\neq c, c=d} m_b m_c^2 \tilde{f}_{bc}^2 - 
\frac{1}{2}
\sum_{b\neq c\neq d}m_{b}m_{c}m_{d} \tilde{f}_{bc}\tilde{f}_{bd}
\eea
and symmetrizing over the summation indices
\bea
V
&=&-\frac{1}{4} \sum_{b\neq c} m_b m_c (m_b + m_c ) \tilde{f}_{bc}^2 \nonumber \\
&& -\frac{1}{6}
\sum_{b\neq c\neq d}m_{b}m_{c}m_{d}\left[\tilde{f}_{bc}\tilde{f}_{bd}
+\tilde{f}_{cb}\tilde{f}_{cd}+\tilde{f}_{db}\tilde{f}_{dc}\right]
\label{Vsymm}
\eea

The term in the last bracket is, in general, a three-body term. We demand that it be, instead, a sum of
two-body terms. That is, we impose the condition
\bea
\tilde{f}_{bc}\tilde{f}_{bd}
+\tilde{f}_{cb}\tilde{f}_{cd}+\tilde{f}_{db}\tilde{f}_{dc}= g_{bc} + g_{bd} + g_{cd}
\eea
for some functions $g_{ab} (x)$, for all distinct $b$, $c$, $d$.

The above is a functional equation similar to equations encountered in the study of the Lax pair of identical Calogero
particles \cite{calfunc}  
%{\color{blue}{M. Bruschi and F. Calogero: General analytic solution of certain functional equations of addition type, SIAM J. Math. Anal. 21(1990), 1019-1030}}, but generalized for distinguishable particles, for which the functions
${\tilde f}_{ab}$ and $g_{ab}$ can depend on the particle indices. Its solution can be obtained with methods
similar to the ones for the indistinguishable case, and is derived in Appendix A. In general, the solution
involves elliptic Weierstrass functions. In this paper we will focus on the simpler case where the functions
$g_{ab}$ are constants, that is
\bea
\tilde{f}_{cb}\tilde{f}_{db}
+\tilde{f}_{dc}\tilde{f}_{bc}+\tilde{f}_{bd}\tilde{f}_{cd}=C_{bcd}
\label{functf}
\eea
where $C_{bcd}$ is a constant independent of $x$. In this case, the three-body term in the potential (\ref{Vsymm}) 
becomes an irrelevant constant and the potential becomes a sum of two-body terms $V_{ab}$.
Up to rescalings of the coordinate $x$, we have the folowing three possibilities for
$\tilde{f}_{ab}(x_{ab})$,  $F_{ab}(x_{ab})$ and $V_{ab} (x_{ab})$ for various values of $C_{bcd}$, described in Table \ref{tab1} with $g$ a constant.

\begin{table}
\caption {Solutions to functional equations without external potential} \label{tab1} 
\begin{center}
\begin{tabular}{|c|c|c|c|}
\hline 
$C_{abc}$ & $\tilde{f}_{ab}(x_{ab})$ & $F_{ab}(x_{ab})$ & $- V_{ab} ( x_{ab} )$ \tabularnewline
\hline 
\hline 
0 & $g/{x_{ab}}$ & $g \, m_a m_b \log\left|x_{ab}\right|$ &{\large  ${g \,m_a m_b (m_a + m_b ) \over 4 \, x_{ab}^2}$}\tabularnewline
\hline 
$-g^2$ & $g \,{\cot x_{ab}}$ & $g \, m_a m_b \log\left|\sin x_{ab}\right|$ & {\large ${g^2 m_a m_b (m_a + m_b ) \over 4 \sin^2 x_{ab}}$}\tabularnewline
\hline 
$+g^2$ & $g\,{\coth x_{ab}}$ & $g \, m_a m_b \log\left|\sinh x_{ab}\right|$ &{\large ${g^2 m_a m_b (m_a + m_b ) \over 4 \sinh^2 x_{ab}}$}\tabularnewline
\hline 
\end{tabular}
\end{center}
\end{table}

The above are essentially the rational, periodic and hyperbolic integrable models of
Calogero type but with particle-dependent two-body couplings. The integrablity of this generalized version 
of family of Calogero models (unequal coupling constants and masses) remains unexplored. 

\subsection{The case $W_a (x ) \neq 0$  (External Potentials)}

We now consider the case where the prepotential includes one-body terms $W_a (x_a )$, leading to
a term $w_a = W'_a (x_a )$ in (\ref{derphi}). In analogy with $f_{ab}$ we define
\be
w_a = m_a {\tilde w}_a
\ee
The expression for the potential, using ${\tilde f}_{ab} = - {\tilde f}_{ba}$ and symmetrizing in the indices,
 becomes
\bea
-V
&=&\frac{1}{4} \sum_{b\neq c} m_b m_c (m_b + m_c ) \tilde{f}_{bc}^2 \nonumber \\
&+& \frac{1}{6}
\sum_{b\neq c\neq d}m_{b}m_{c}m_{d}\left[\tilde{f}_{bc}\tilde{f}_{bd}
+\tilde{f}_{cb}\tilde{f}_{cd}+\tilde{f}_{db}\tilde{f}_{dc}\right] \nonumber \\
&+& \half \sum_{b\neq c} m_b m_c ({\tilde w}_b - {\tilde w}_c ) {\tilde f}_{bc} + \half \sum_b m_b {\tilde w}_b^2
\label{VWsymm}
\eea
The terms in the first two lines are the same as for the $W_a = 0$ case, and the requirement that they reduce to 
two-body terms gives the same solutions for ${\tilde f}_{ab}$ as before. The last line contains two-body and one-body
potentials. Demanding that two-body terms depend only on particle distance imposes the condition
\be
 ({\tilde w}_b - {\tilde w}_c ) {\tilde f}_{bc} = u_{bc} + v_b + v_c
\label{bc}
\ee
for some particle-dependent functions $u_{ab} (x)$ and $v_a (x)$.
The above is a functional equation for the functions ${\tilde w}_a$ whose solutions depend on the functions
$ {\tilde f}_{ab}$. Its treatment is given in Appendix B, and we state in Table \ref{tab2} the solutions in each case,
up to rescalings of $x$, with $m_{tot} = \sum_a m_a$ the total mass, $C_a, c_1 , c_2 , c_3$ constants, and as usual $C_{ab} = C_a - C_b$.

\begin{table}
\caption {Solutions to functional equations with external potential} \label{tab2} 
\begin{center}
{\hspace*{-1.1cm}
%\begin{tabular}{|c|c|c|c|}
%\begin{tabular}{|c|c|c| c{5cm}|}
\begin{tabular}{|p{1.5cm}|p{2.9cm}|p{2.5cm}|p{8.4cm}|}
\hline   
$\tilde{f}_{ab}(x_{ab})$ & $~~~~{\tilde w}_a ( x_a )$ & $~~~u_{ab} ( x_{ab} )$ & $~~~~~~~~~~~~~~~~~
-2 V_a (x_a )$\tabularnewline
\hline 
\hline 
{\vspace*{-0.1cm}$g/{x_{ab}}$ }& $C_a + c_1 x_a + c_2 \, x_a^2 + c_3 \, x_a^3$ & { {\vspace*{-0.15cm}{\large$\frac{g C_{ab}}{x_{ab}}$}$-\half{g c_3} x_{ab}^2$}}& 
{\vspace*{-0.1cm} $m_a {\tilde w}_a^2 +
g (m_{tot}-m_a) m_a (c_2 \, x_a + {3 \over 2} c_3 \, x_a^2)$}\tabularnewline
\hline 
{\vspace*{-0.1cm}$g \cot x_{ab}$}& $C_a + c_1 \cos 2x_a + c_2 \sin 2x_a + c_3 x_a $&{$g C_{ab}\cot x_{bc}+g c_3 x_{ab} \cot x_{ab}$ }& \vspace*{-0.1cm}{$m_a {\tilde w}_a^2 + g
(m_{tot}-m_a) m_a ( c_2 \cos 2x_a - c_1 \sin 2x_a) $}\tabularnewline
\hline 
{\vspace*{-0.1cm}$g \coth x_{ab}$} & \small{$C_a + c_1 \cosh 2x_a + c_2 \sinh 2x_a + c_3 x_a$}&$g C_{ab} \coth x_{bc}+g c_3 x_{ab} \coth x_{ab}$ &{\vspace*{-0.1cm} $m_a {\tilde w}_a^2 + g
(m_{tot}-m_a) m_a (c_2 \cosh 2x_a + c_1 \sinh 2x_a )$}\tabularnewline
\hline 
\end{tabular}
}
\end{center}
\end{table}

In the case that we restrict our solutions to $u_{bc} = 0$ (a justification of why this may be relevant is given later), that is, we only allow one-body terms to appear
in the right hand side of (\ref{bc}), then $c_3$ must vanish and all $C_a$ have to be equal. The acceptable forms for ${\tilde w}_a$ and coresponding one-body potentials are given in Table \ref{tab3}. Interestingly, we recover the restricted form of the integrable potentials found in \cite{i3,APnew1, APnew2}, for which the proof of integrability simplifies considerably. (The most general class of integrable potentials, derived in \cite{APnew1,APnew2},
depends on one additional parameter.)
\begin{table}
\caption {Solutions to functional equations with one-body external potentials} \label{tab3} 
\begin{center}
{\hspace*{-1cm}
\begin{tabular}{|c|c|c|}
\hline 
$\tilde{f}_{ab}(x_{ab})$ & ${\tilde w}_a ( x_a )$ & $-2 V_a (x_a )$\tabularnewline
\hline 
\hline 
$g/{x_{ab}}$ & $c_0 + c_1 x_a + c_2 x_a^2$ & { $m_a {\tilde w}_a^2 + g
(m_{tot}-m_a) m_a c_2 \, x_a$}\tabularnewline
\hline 
{$g \cot x_{ab}$}&{$c_0 + c_1 \cos 2x_a + c_2 \sin 2x_a$ }&{ $m_a {\tilde w}_a^2 + g
(m_{tot}-m_a) m_a ( c_2 \cos 2x_a - c_1 \sin 2x_a ) $}\tabularnewline
\hline 
{$g \coth x_{ab}$} & \small{$c_0 + c_1 \cosh 2x_a + c_2 \sinh 2x_a$} &{ $m_a{\tilde w}_a^2 +
g (m_{tot}-m_a) m_a ( c_2 \cosh 2x_a + c_1 \sinh 2x_a )$ }\tabularnewline
\hline 
\end{tabular}
}
\end{center}
\end{table}

It would seem peculiar that the eliminated term is the cubic term in the rational case, while it is the linear term in the
trigonometric and hyperbolic cases. We can check, however, that the small-$x$ limit of the above trigonometric or hyperbolic
potentials, upon proper scaling
of the coefficients, reduces to the rational case. In particular, the presence of the linear term
in the trigonometric and hyperbolic cases  introduces an extra parameter which can be tuned to make 
the coefficient of $x^3$ finite and nonzero in the small-$x$ limit.

\subsection{Reality Conditions and Solitons}

The potential of the above dynamical system, as given by (\ref{V}), is negative definite and thus, in principle,
unstable. To obtain more interesting stable systems, we extend the variables and parameters to the complex
plane. The goal is to find a set of parameters for which the particles, or at least a subset of them, will
remain on the real axis and will constitute a stable real dynamical system.

With an appropriate choice of parameters, we can ensure that the potential will  turn positive and thus be
stable when the $x_a$ are on the real axis. This is simply achieved by taking $\Phi \to i \Phi$, that is,
\be
{\tilde f}_{ab} \to i {\tilde f}_{ab} ~,~~{\tilde w}_a \to i {\tilde w}_a ~,~~ V \to -V
\ee
The first-order equations of motion in terms of $i {\tilde f}_{ab}$ and $i {\tilde w}_a$, however, become
\be
{\dot x}_a = i \sum_{b \neq a} m_b {\tilde f}_{ab} + i {\tilde w}_a
\label{xadot}
\ee
We observe that even if the initial positions of the particles are real, their velocities will be imaginary and
therefore will escape into the complex plane. It is thus impossible to obtain a stable system with
all the particle coordinates real.

We demand, therefore, that a subset of the particle coordinates, say $x_1 , \dots x_N$, to remain real,
while the rest of them, $x_{N+1} , \dots x_n$, will become complex. For clarity, we call the real coordinates
$x_j$, $j=1,\dots N$, and the remaining complex ones $z_\alpha$, $\alpha = 1, \dots M$ with $M=n-N$.
For reasons that will become apparent later on, we call the $z_\alpha$ ``solitons".

The first basic requirement is that the potential should contain no couplings between $x_j$ and $z_\alpha$,
else such mutual forces would drive the $x_j$ into the complex plane. This means, in particular, that the
top line in (\ref{VWsymm}) should contain no mixed terms; that is
\be
m_j m_\alpha (m_j + m_\alpha ) = 0 ~~{\rm for~all}~~j,\alpha
\ee
The only possibility (other than $m_j =0$ or $m_\alpha =0$) is
\be
m_j = - m_\alpha = m
\ee
That is, all particles have the same (positive) mass, while solitons have the same negative mass. The second
line in (\ref{VWsymm}) is a constant in our case, so it  is of no concern. The first term in the third line, however,
in principle couples particles and solitons, as its mixed terms are
\be
m_j m_\alpha u_{j\alpha}
\ee
We thus demand $u_{j\alpha} = 0$, which leaves the possibilities found before for vanishing $u_{ab}$.
The above conditions will secure that the potential does not couple (real) particles and (complex) solitons
and thus their second-order equations of motion decouple.
Particles will obey their own Calogero-like equation of motion and solitons will
obey their own similar equations.

To make sure that the motion of particles will remain real, we must further ensure that their initial velocities are
real. Their expression (for $m_j =m$ and $m_\alpha = -m$) is
\be
{\dot x}_j = i m \sum_{k (\neq j)} {\tilde f}_{jk} - i m \sum_{\alpha} {\tilde f}_{j\alpha} + i {\tilde w}_j
\label{xri}
\ee
while the corresponding expression for solitons is 
\be
{\dot z}_{\alpha} = -  i m \sum_{k } {\tilde f}_{\alpha k} + i m \sum_{\beta (\neq \alpha)} {\tilde f}_{\alpha \beta} + i {\tilde w}_{\alpha}
\label{zjs2}
\ee
For real $x_j$, the first and last terms in the right hand side of (\ref{xri}) are purely imaginary. Reality of the ${\dot x}_j$
implies
\bea
&& m \sum_{k (\neq j)} {\tilde f}_{jk} - m~ {\rm Re} \left( \sum_{\alpha} {\tilde f}_{j\alpha} \right) 
+ {\tilde w}_j = 0 \label{Imz} \\
&&{\dot x}_j = m ~{\rm Im} \left(\sum_{\alpha} {\tilde f}_{j\alpha} \right)
\label{Rez}
\eea
The above are, altogether, $2N$ real constraints for $2N+ 2M$ real variables (the $N$ initial positions $x_j$,
$N$ initial velocities ${\dot x}_j$ and the $M$ complex initial values of complex $z_\alpha$). Overall, there are
$2M$ free parameters in this model, which can be chosen as the initial values of $z_\alpha$. (We caution that
the parameters $z_\alpha$
may not actually uniquely determine the state of the system, as equation (\ref{Imz}) may have none or more than one
solutions for the $x_j$.)

We see that, in general, the real particle system $x_j$ follows a motion that is parametrized by the initial values
of the $M$ complex soliton parameters $z_\alpha$, which can, then, be viewed as phase space variables for
the system. The number of solitons $M$ determines the degrees of freedom of the system. In particular,
for $M=0$ all velocities are zero and particles are in their equilibrium position determined by (\ref{Imz})
for ${\tilde f}_{j\alpha} =0$.

For the case $M=1$, $z_1 (t)$ moves as a solitary particle inside the one-body potential $V (z_1 )$
given in Table 3. Its presence deforms the distribution of $x_j$, creating a coalescence of
particles near Re$(z_1 (t) )$, that is, a solitary wave. The imaginary part of $z_1 (t)$ determines the
width of the wave, while also determining the velocity of the soliton. For $M>1$ the various $z_\alpha$
move as particles in the potential $V ( z)$ interacting through Calogero-type potentials, while the real
particles $x_j$ perform a motion ``guided" by the $z_\alpha$. The above features justify the identification
of $z_\alpha$ as soliton parameters, and the corresponding motion of $x_j$ as solitonic many-body ``waves".

\subsection{Stability}

In the previous section we ensured that the dynamical system described by the $x_j$ obeys second-order equations
of motion in a stable potential. We still need to ensure, however, that the initial value problem as determined
by the choice of soliton parameters $z_\alpha$ has solutions. As we shall see, this imposes nontrivial constraints.

We start with the simplest case of no solitons. As stated, this corresponds to static $x_j$ obeying the condition
\be
m \sum_{k (\neq j)} {\tilde f}_{jk} + {\tilde w}_j = 0
\ee
or, using the definitions of ${\tilde f}_{jk}$ and ${\tilde w}_j$ as well as $m_j = m_k = m$,
\be
{\partial \Phi \over\partial x_j} = 0 ~,~~~ \Phi = \sum_{j<k} F_{jk} (x_{jk}) + \sum_j W(x_j )
\ee
These are the equilibrium conditions for particles $x_j$ inside the (real) prepotential $\Phi$ defined above. In order
for this to have solutions it must be of an appropriate form.

Let us examine the rational case. The prepotential and real potential in this case become
\bea
\Phi = \sum_j \left( c_0 x_j + {c_1 \over 2} x_j^2 + {c_2 \over 3} x_j^3 \right) + \sum_{j<k} g \, m \ln | x_{jk} |
~~~~~~~~~~\cr
V= \sum_j \left[ {m \over 2} \left( c_0  + c_1  x_j + {c_2} x_j^2 \right)^2 + g (N-1) m^2 c_2 x_j
\right] + \sum_{j<k} {g \, m^2 \over x_{ab}^2}
\label{Vquar}
\eea
The prepotential consists of a {\it cubic} one-body potential and a logarithmic two-body interaction. We can always 
choose the sign
of $g$ that makes the interaction term repulsive, $g <0$ (in the opposite case we can flip the sign of
$g$ as well as $c_0 , c_1 ,  c_2$ which leads to the same potential $V$). 
So this becomes a standard problem of finding the
equilibrium position of particles in a potential with repulsive logarithmic interactions.

Generically, this problem may not have solutions since the cubic potential is unbounded from below. This happens,
in particular, for the simplest nontrivial purely cubic case ($c_0 = c_1 = 0$). To guarantee the existence of solutions
the prepotential must have a ``well" such that it can trap particles. The existence of such a well
requires the condition
\be
c_1^2 > 4 c_0 c_2
\ee
This condition is necessary but not sufficient. The well must also be deep enought to hold $N$ particles repelling
each other with a logarithmic interaction of strength $g m$. We have no explicit expressions for this restriction in
terms of the constants of the problem. We stress that the above restrictions are necessary so that the equilibrum
problem can be addressed in the first-order formalism. The real potential $V$, being a quartic expression in the coordinate,
always has an equilibrium solution.

For a nonzero number of solitons the situation becomes more complicated, as now the set of equations
(\ref{Imz}, \ref{Rez}) must admit solutions. In general this will imply restrictions to both the form of the
potential and the range of ``phase space" parameters $z_\alpha$. We point out, however, that the interaction
between particles and solitons is attractive, so in general the presence of solitons improves the situation.
In fact, it may be that the unstable prepotentials for which the zero-soliton case has no static solutions
become stable in the presence of a large number of solitons, akin to a ``broken" symmetry phase with a soliton condensate. We postpone the investigation of this issue for a future publication.

\subsection{A Note on Integrability}

Solitons are a hallmark of integrability and, indeed, Calogero-like systems are one of the most celebrated
classes of integrable models. The systems derived here, however, differ from standard Calogero models in the fact that
(i) the particles are not identical, with distinct masses and corresponding two-body interactions, and (ii) in the presence
of a more general one-body potential.

In the above analysis, the issue of integrability remains unaddressed. There are, nevertheless, some intriguing
indications that these systems are integrable. First, the reality conditions necessary to have stable potentials
naturally restrict the system to identical particles ($m_j = m$) which is integrable. Further, the extended
one-body potentials that we obtained (quartic in the rational case, harmonic with two nontrivial modes in the
trigonometric and hyperbolic cases) are exactly the potentials that are know to be integrable \cite{APnew1,APnew2}.

In fact, the obtained potentials belong to a somewhat restricted class.
Specifically, a generic quartic potential has 4 nontrivial parameters (ignoring the constant term). Our potential,
however, has only 3 parameters ($c_0, c_1 , c_2$), being essentially a complete square. Although arbitrary quartic
potentials are integrable, for the above restricted potentials of ``Bogomolny" type the proof of integrability simplifies
considerably \cite{APnew1, APnew2, i3}.

Finally, an inspection of the known integrals of motion for the standard quadratic Calogero model ($c_3 = 0$)
reveals that they are all zero. (This is not a contradiction, since particles and solitons contribute with opposite
signs.) Although this is not yet a proof of integrability, since the value of higher integrals could fluctuate
between particles and solitons, it is a tantalizing clue that a general proof of integrability may well be
possible within the first-order formalism. This and other issues are saved for future investigation.

%%%%%%%%%%%%%%%%%%%%%%%%%%%%%%%%%%%%%%%%
\section{Dual representation and solitons of the rational model in external quartic potentials }
 \la{dualCM}
%%%%%%%%%%%%%%%%%%%%%%%%%%%%%%%%%%%%%%%%

In this section we focus on the particular concrete example of rational Calogero models in external quartic potentials.  Using the general formalism of the previous section, we demonstrate the existence of a dual version involving soliton variables $z_\alpha$ and derive analytical and numerical solutions.

%The system (\ref{csmeq1},\ref{csmeq2}) is rewritten as equations symmetric in $x_{j}$ and $z_{j}$ (see (\ref{xjdot},\ref{zjdot}) below). We refer to the obtained symmetric system as a self-dual form.
%The self-dual form (\ref{xjdot},\ref{zjdot}) makes explicit the relationship between particles $x_{j}$ and excitations (parametrized by $z_{j}$) of Calogero system in external potential. 
%After establishing self-dual form  we consider different reductions of this system. Note that reductions of the number of points $z_{j}$  in a dual model in conjugation with a real reduction ($x_{j}$ - real) produce M-soliton solutions for the original model.
%

%%%%%%%%%%%%%%%%%%%
\subsection{First order equations for the rational Calogero case}
 \la{sec:dcm}

The  general formalism discussed above greatly helps in writing down the first order equations for the rational Calogero case.  If we take the first row of Table 2 and impose that particles (indexed by $j$) have mass $m_j =1$ and solitons (indexed by $\alpha$) have mass $m_\alpha =-1$, then (\ref{xri}) and  (\ref{zjs2})  give
%
%To begin, let us consider $x_{j},z_{j}$ as well as $p_{j}=\dot{x}_{j}$ and $\dot{z}_{j}$ as complex numbers.
%We introduce the following dynamic system:
\bea
	\dot{x}_{j} -iw(x_{j}) & = & -ig\sum_{k=1 (k\neq j)}^{N}\frac{1}{x_{j}-x_{k}}
	+ig\sum_{\alpha=1}^{M}\frac{1}{x_{j}-z_{\alpha}},
 \la{xjdot} \\
   	\dot{z}_{\alpha} -iw(z_{\alpha})& = & ig\sum_{\alpha=1(\alpha \neq \beta)}^{M}\frac{1}{z_{\alpha}-z_{\beta}}
	-ig\sum_{j=1}^{N}\frac{1}{z_{\alpha}-x_{j}},
 \la{zjdot}
\eea
for $x_{j}(t)$ with $j=1,2,\ldots, N$ and $z_{\alpha}(t)$ with $\alpha=1,2,\ldots,M$. Here, the function $w$ is given in Table 3 as
\be
w (x) = c_0 + c_1 x + c_2 x^2
\ee
Since both $w_a$ and $V_a$ are independent of the particle index $a$ (see Table 3), we dropped the subscripts.

Equations (\ref{xjdot},\ref{zjdot}) are first order in time and, for complex coordinates, the dynamics is fully defined by the initial values of $x_{j}$, $z_{\alpha}$, i.e., by $N+M$ complex numbers ($2N+2M$ real variables). If the particle coordinates $x_j$ are real (with real velocities), as discussed in section 2.3, the dynamics is fully determined by the $M$ complex initial values of
$z_\alpha$. Applying Eq. (\ref{V}) of the general formalism, the corresponding second order equations are given by 
\bea
   \ddot{x}_{j} & = & -\frac{g^{2}}{2}\frac{\partial}{\partial x_{i}}\sum_{i\neq j}^{N}
   \frac{1}{\left(x_{i}-x_{j}\right)^{2}}-V^{\prime}(x_{j}), \qquad j=1,\dots, N
 \la{pmotx}
 \\
    \ddot{z}_{\alpha} & = & -\frac{g^{2}}{2}\frac{\partial}{\partial z_{\alpha}}\sum_{\alpha=1 (\alpha \neq \beta)}^{M}\frac{1}{\left(z_{\alpha}-z_{\beta}\right)^{2}}-V^{\prime}(z_{\alpha}), \qquad \alpha=1,\dots, M
 \la{pmotz}
\eea
%The system is a complex version of  the system of equations of motion obtained from hCM (\ref{hCM}), i.e., equivalent to (\ref{csmeq1},\ref{csmeq2}). 
We refer to the system (\ref{xjdot},\ref{zjdot}) as the \textit{dual Calogero system in external potential $V$}. We emphasize that the general formalism greatly simplifies the transition from first order to second order equations. Lack of such a general formalism would have involved a laborious algebra to arrive at the second order equations (\ref{pmotx},\ref{pmotz}). We also point out that the derivation of (\ref{pmotx},\ref{pmotz}) from (\ref{xjdot},\ref{zjdot}) holds for
arbitrary $N$ and $M$, although we are mostly interested in the case $M<N$.

\subsection{Multi-Soliton solutions}
 \la{sec:Msolutions}
%%%%%%%%%%%%%%%%%%%%%%%%%%%%%%%%%%%%%%%%
From the first order equations derived above for the rational Calogero model, we get (separating real and imaginary parts of (\ref{xjdot})), 
\bea
	w(x_{j}) &=& g\sum_{k=1 (k\neq j)}^{N}\frac{1}{x_{j}-x_{k}}
	-\frac{g}{2}\sum_{\alpha=1}^{M}\left(\frac{1}{x_{j}-z_{\alpha}}+\frac{1}{x_{j}-\bar{z}_{\alpha}}\right),
 \la{xjreal} \\
	{\dot x}_j = p_{j}& =&
	i\frac{g}{2}\,\sum_{\alpha=1}^{M}\left(\frac{1}{x_{j}-z_{\alpha}}-\frac{1}{x_{j}-\bar{z}_{\alpha}}\right).
 \la{pjreal}
\eea
As we noted in section 2.3, if we specify the $M$ complex positions $z_{\alpha}$ at any time we can find both the $N$ real positions $x_{j}$ and the corresponding real momenta $p_{j}$. If we are given the initial values of $x_{j}$ and $p_j$, their evolution is fully determined by (\ref{pmotx}). However, these initial values are not, in general, independent, as they are related by (\ref{xjreal},\ref{pjreal}) through the values of the $M$ complex parameters ($2M$ real parameters) $z_{\alpha}$ . (Only when $M \ge N$ we can choose them independently.) The initial values of ${\dot z}_\alpha$
are always restricted, as they need to satisfy (\ref{zjdot}) and the $x_j$ are fully fixed by the $z_\alpha$.

%The function $Real(\Phi)$ coincides with an ``electrostatic energy'' of $N$ particles with unit charges interacting through a logarithmic potential (2d Coulomb potential). 
%Very importantly, these particles are restricted to move along a straight line (a real axis) and are in the presence of $2M$ external charges $-1/2$ placed at $z_{n},\bar{z}_{n}$ and an external potential  $W(x)$.
%The solution of (\ref{xjreal}) is not necessarily a minimum of (\ref{estatic}). Soliton solutions correspond to any extremum (maximum, minimum or saddle point) of (\ref{estatic}).

%It is important to stress that here and in the following we choose the signs $\omega>0$ and $g>0$ which guarantees that the harmonic potential in (\ref{estatic}) is confining.

%%%%%%%%%%%%%%%%%%%
\subsection{Solution for zero solitons}
 \la{background}

In this section we discuss the case of zero solitons. This gives rise to a spacially inhomogeneous static background configuration where all particles are at rest. (We also note that this is the same as the limit where all $z_\alpha$ 
go to infinity.) Taking $M=0$ we have $p_{j}=0$ for all $j$ and the particle coordinates settle in the equilibrium
positions
\be
\la{Hermite}
	w (x _j ) = c_0 + c_1 x_j + c_2 x_j^2 = g \sum_{k=1 (k\neq j)}^{N} \frac{1}{x_j-x_k}.
\ee
If we just had a harmonic trap as in \cite{kul1,kul2}, in which case $w^{harm}(x)=\omega x$,
then it is known that the solution of this system of algebraic equations is given by the roots of the $N$-th Hermite polynomial (Stiltjes formula \cite{Szego-1975,mehta}). We are unaware of a generalization of the Stiltjes formula when $w(x)$ has a quadratic form as above.

%%%%%%%%%%%%%%%%%%%
\subsection{The single-soliton solution}
 \la{1soliton}

The single soliton case is essentially equivalent to one complex $z$ coordinate moving freely in a quartic polynomial potential.  For $M=1$ equations (\ref{xjreal},\ref{pjreal}) become
\bea
	c_0 &+& c_1 x_j + c_2 x_j^2 = g\sum_{k=1 (k\neq j)}^{N} \frac{1}{x_j-x_k}
	-\frac{g}{2} \left( \frac{1}{x_j -z} + \frac{1}{x_j-\bar{z}}\right),
 \la{M1xj} \\
   	p_j &=&  i\frac{g}{2} \left( \frac{1}{x_j-z} - \frac{1}{x_j-\bar{z}}\right).
 \la{M1pj}
\eea
Equation (\ref{M1xj}) is a further generalization of the Stieltjes problem (\ref{Hermite}) (see Refs. \cite{Szego-1975,forrester,orive}). To our understanding, exact solutions of (\ref{M1xj}) are not known (not even in the case of harmonic potential), and the solution may not be unique. Equation (\ref{M1xj}) is essentially the equilibrium position of $N$ particles repelling each other but held in an external potential and also attracted to an additional particle of opposite ``charge", as will be further elaborated in section 3.6.

The soliton is a single particle moving in an external quartic polynomial potential. That is,
equation (\ref{pmotz}) in the case $M=1$ takes the simple form
\be
\ddot{z} = -V^{\prime}(z) = -k_3 z^3 - k_2 z^2 - k_1 z - k_0 
 \la{oscillator}
\ee
where $k_0 , \dots , k_3$ are related to the parameters of the model (see Table 3 and eq.\ (\ref{Vquar})).
This is a single anharmonic complex oscillator. Typically, analytical solutions for the above equation are not available for quartic polynomials $V(z)$, but solving a single particle problem is numerically easy. Knowing the value of $z(t)$, we then use (\ref{M1xj}) and (\ref{M1pj}) to find the $x_j$ and $p_j$. The upper left panel of Fig.\ 1 demonstrates the 
particle evolution profile for the case of a single solition. 
We see that the worldline of particles clearly shows a robust soliton. The motion of the corresponding variable $z(t)$ is given in the lower left panel of the same Fig.\ 1.

\subsection{The two-soliton solution}

The case when the system has two solitons is nontrivial even with respect to the $z_1 , z_2$ variables. In this case, we have two complex soliton coordinates in a quartic potential interecting with Calogero forces. Putingt $M=2$ in (\ref{xjreal},\ref{pjreal}) we
obtain
\bea
	w(x_{j}) &=& g\sum_{k=1 (k\neq j)}^{N}\frac{1}{x_{j}-x_{k}}
	-\frac{g}{2}\Big(\frac{1}{x_{j}-z_{1}}+\frac{1}{x_{j}-\bar{z}_{1}} \la{xjreal2}\\
 &+& \frac{1}{x_{j}-z_{2}}+\frac{1}{x_{j}-\bar{z}_{2}}\Big),
 \nonumber\\
	p_{j}& =&
	i\frac{g}{2}\left(\frac{1}{x_{j}-z_{1}}-\frac{1}{x_{j}-\bar{z}_{1}} + \frac{1}{x_{j}-z_{2}}-\frac{1}{x_{j}-\bar{z}_{2}} \right)
 \la{pjreal2}
\eea
while solitons obey the coupled second-order equations
\bea
\ddot{z}_{1} & = & -\frac{g^{2}}{2}\frac{\partial}{\partial z_{1}}\sum_{i\neq j}^{M}\frac{1}{\left(z_{1}-z_{2}\right)^{2}}-V^{\prime}(z_{1}) \\
\ddot{z}_{2} & = & -\frac{g^{2}}{2}\frac{\partial}{\partial z_{2}}\sum_{i\neq j}^{M}\frac{1}{\left(z_{2}-z_{1}\right)^{2}}-V^{\prime}(z_{2})
\eea
Initial conditions for ${\dot z}_{1,2}$ can be set by specifying $z_{1,2}$ and using (\ref{xjreal2})
to find $x_j$ and then (\ref{zjdot}) to find ${\dot z}_{1,2}$. Then the evolution of $z_{1,2}$ can be found
by solving the above Calogero equations.

The upper right panel of Fig. 1 shows the 
particle evolution profile in the case of two solitions. 
We see that the worldline of particles clearly shows two robust solitons, one moving left and another moving right. The motion of the corresponding complex soliton variables is given in the lower right panel of the same Fig. 1.

\subsection{Mapping solitons to an electrostatic problem and the numerical protocol}

Let us take a close look at equation (\ref{xjreal}). It is essentially the derivative of the real part of the
prepotential $\Phi$:
\bea
U = Real(\Phi) &=& \sum_{j=1}^{N}W(x_j) -\sum_{j<k}\ln|x_{j}-x_{k}| \nonumber\\&+&\frac{1}{2}\sum_{j=1}^{N}\sum_{n=1}^{M}\Big[\ln|x_{j}-z_{\alpha}|+\ln|x_{j}-\bar{z}_{\alpha}|\Big].
 \la{estatic}
\eea
where we remind the reader that the function $W(x)$ is related to $w(x)$ as 
\bea
W(x)=\int_{0}^{x}w(x^{\prime})dx^{\prime}
\eea
Eq. (\ref{xjreal}) is then the extremum condition ${\partial U \over \partial x_j} = 0 $ for the above function. 

The function $U$ can be thought of as the ``electrostatic energy'' of $N$ particles with unit charges interacting through a logarithmic potential (2d Coulomb potential), restricted to move along a straight line (the real axis) and in the presence of $2M$ external charges $-1/2$ placed at $z_{\alpha},\bar{z}_{\alpha}$ and of an external potential  $W(x)$.
The solution of (\ref{xjreal}) is not necessarily a minimum of (\ref{estatic}), but may correspond to any fixed point (maximum, minimum or saddle point) of (\ref{estatic}), and there may be several such points.

The above observation forms the basis for a numerical procedure for solving  Eq. (\ref{xjreal}), at least for solutions
corresponding to a local minimum. The basic idea is to let the particles slide towards the minimum of the
above potential by introducing a ``viscous" force that allows them
to move towards their equilibrium positions.
That is, we introduce the following $N$ coupled ODEs 
\bea
\dot{x}_j = -\gamma \frac{\partial{U}}{\partial x_j}
\eea
It is clear that the above drives the system to the minimum of the potential $U$. In fact, the above equation
implies 
\bea
\label{ueq}
\frac{d U}{dt}
=-\frac{\gamma}{2}  \sum_{j=1} ^{N}  \left( \frac{\partial{U}}{\partial x_j} \right)^2
\eea
so the potential decreases until it reaches a fixed point. Local maxima can also be dealt this way by flipping
the sign of $U$ and ttuning them into minima. Saddle points, on the other hand, will be missed.

The above first-order equation can be integrated numerically. Once we find the solutions for $x_j$, we then use Eq. (\ref{M1pj}) to find the initial momenta. These form the special initial conditions for the particles that correspond to
a set of solitons, and we can evolve them according to Eq. \ref{pmotx} without further reference to the soliton
variables.Therefore, we have mapped the problem of finding soliton configurations to an electrostatic problem of a function $U$.

\begin{figure}
\includegraphics[scale=0.48]{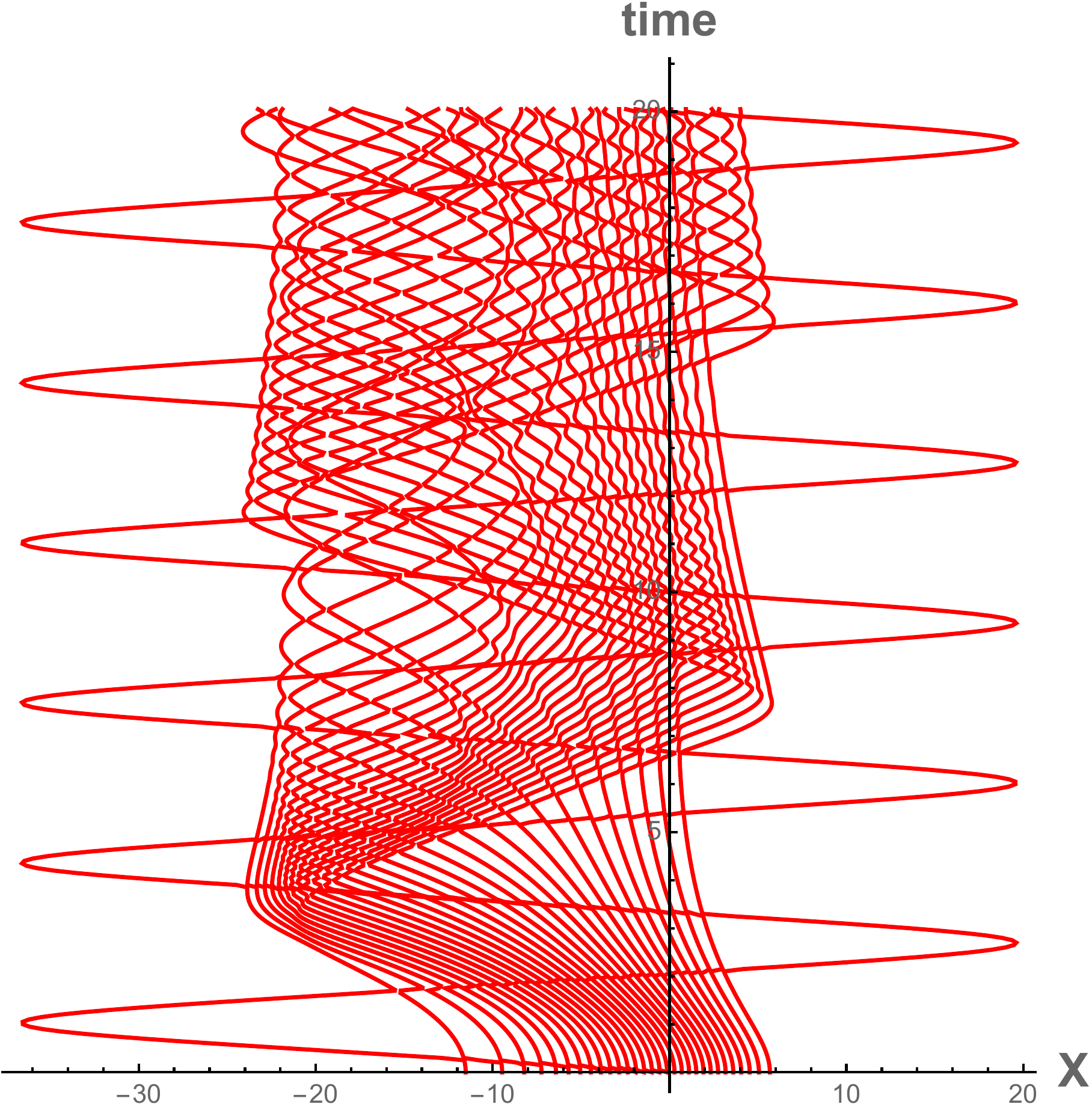}
\includegraphics[scale=0.55]{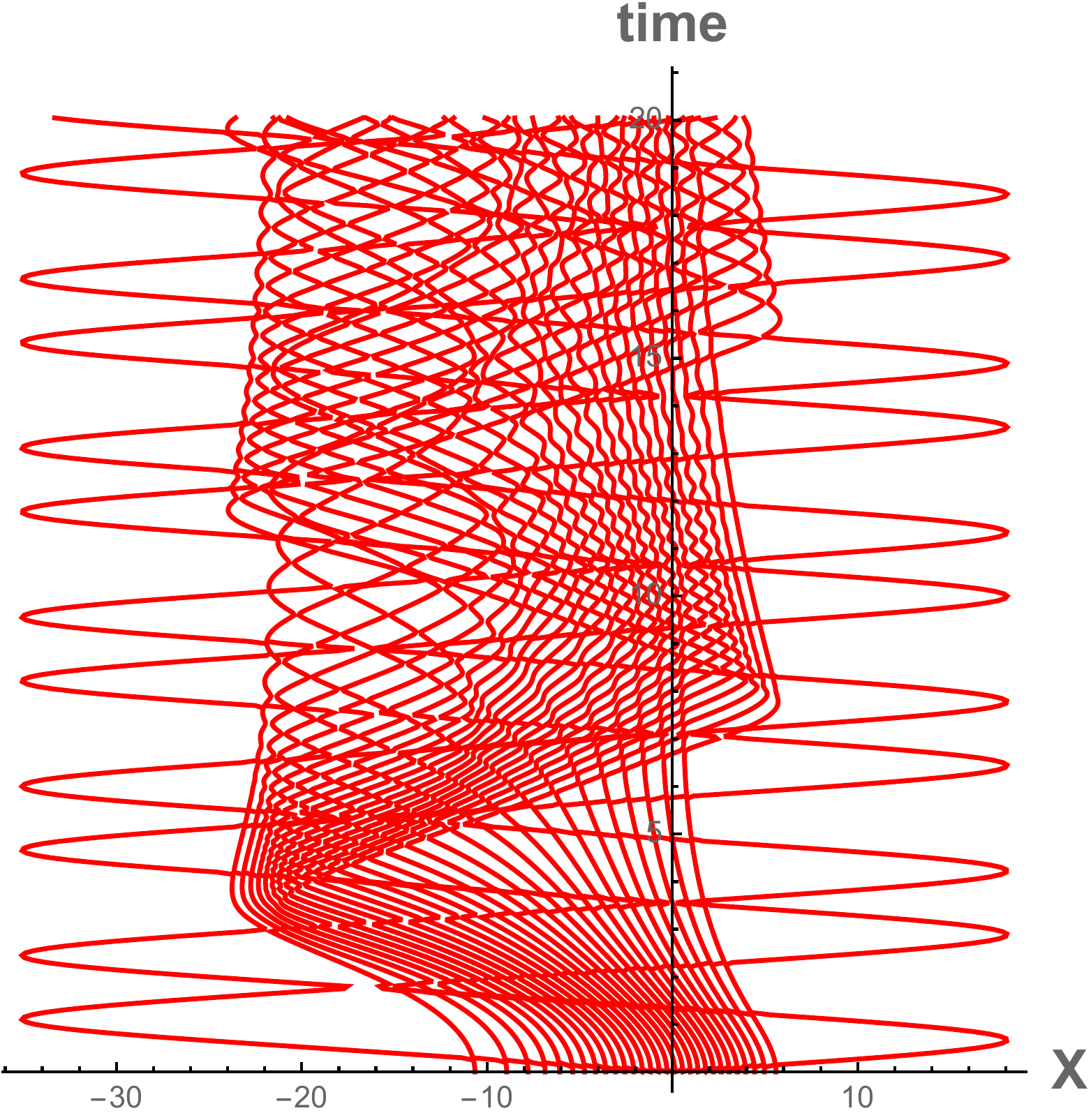}
\includegraphics[scale=0.55]{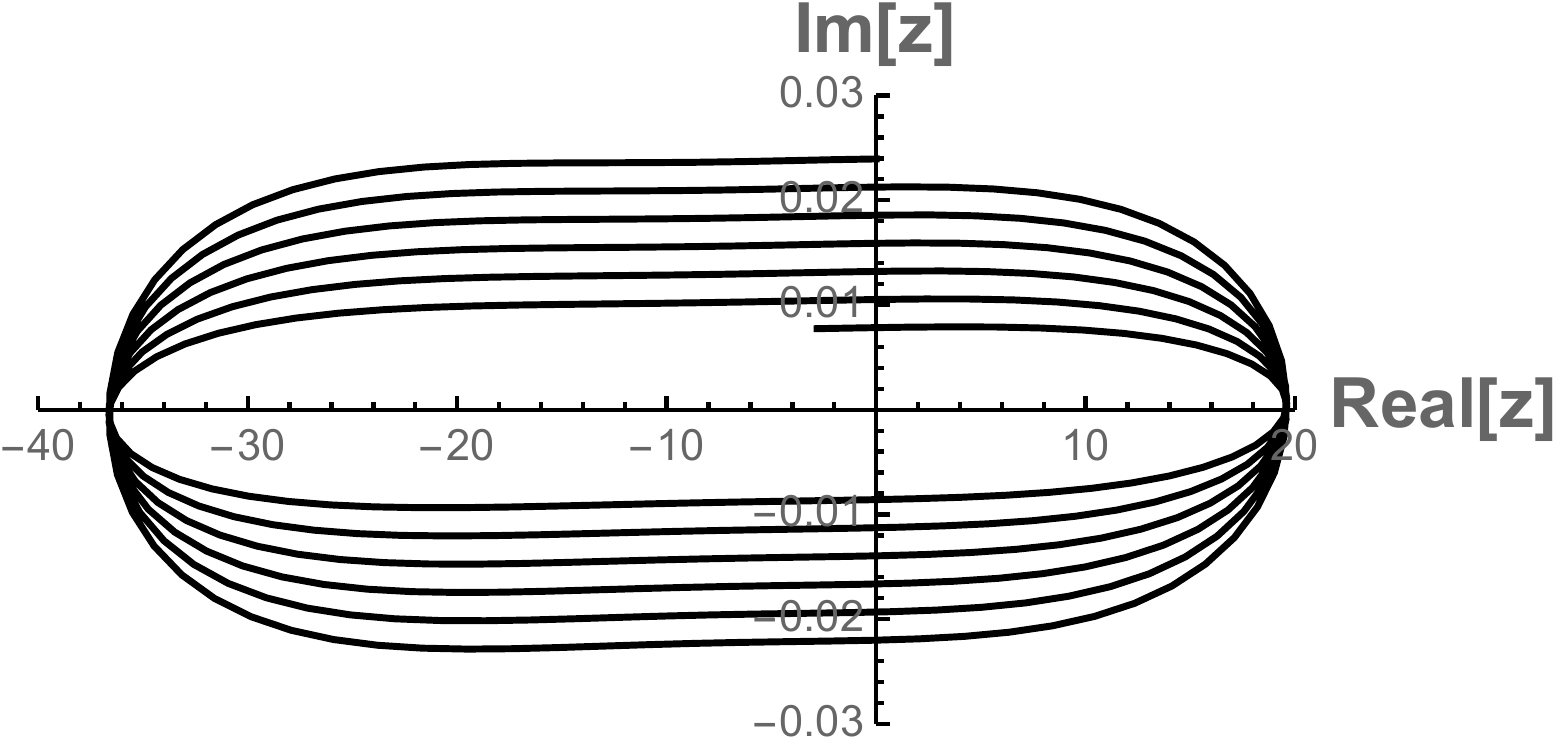}
\includegraphics[scale=0.30]{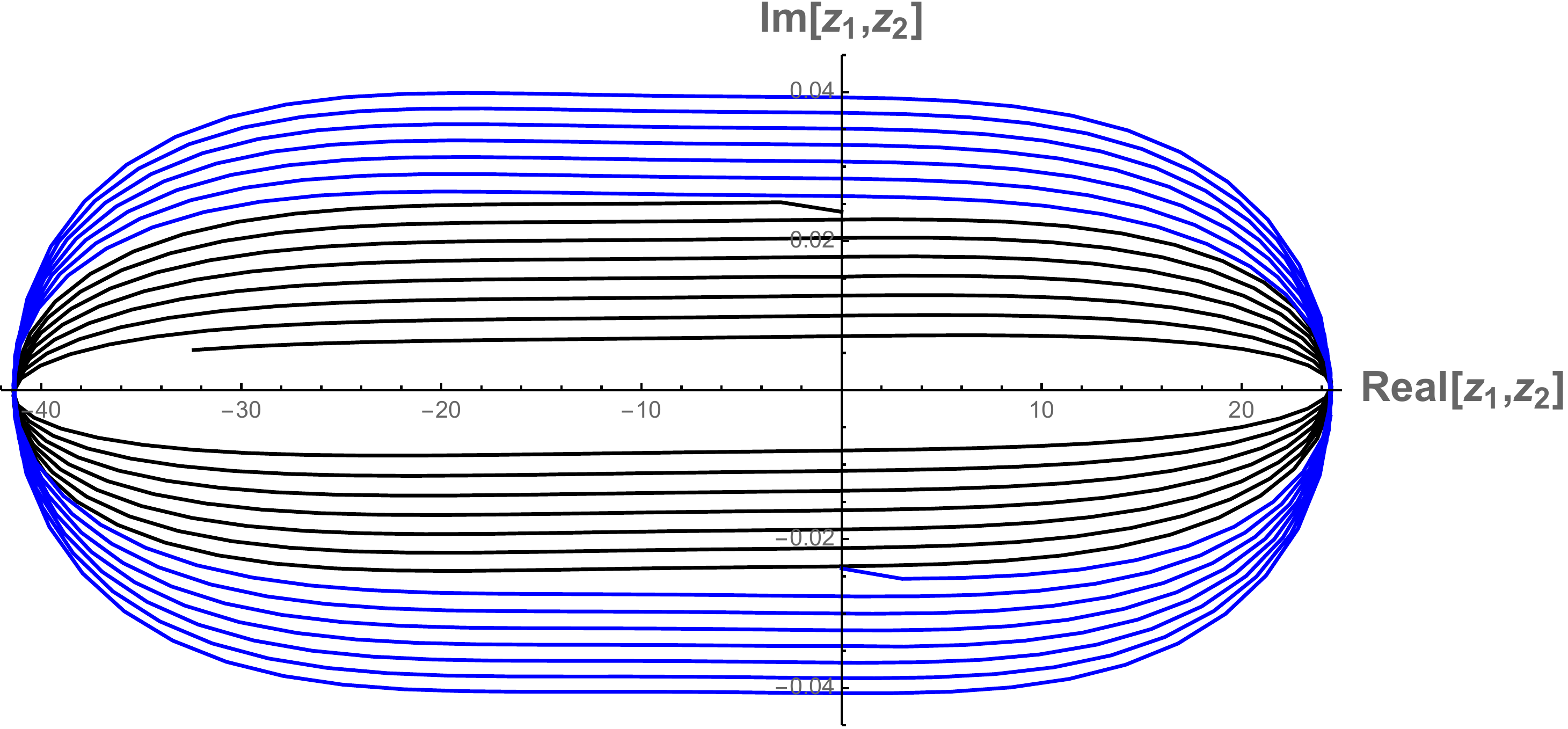}
\caption{\textbf{(Upper Panel Left)} One soliton solution for rational Calogero model in external quartic polynomial potential. Here we take N=31 particles and the special initial condition is dictated by a single $z = 0.0239 i $. We notice, in the worldine picture, a remarkable evidence of a robust soliton We use the prepotential $w(x) = x +0.06 x^2$ which makes the potential
 %$V(x) =30x+x^2 + 0.12 x^3 + 0.0036 x^4$. 
$V(x)= 0.87 x + 0.5 x^2 + 0.06 x^3 + 0.0018 x^4$ using first row of Table 3. 
\textbf{(Lower Panel Left)} We show the motion of one dual variable ``z" corresponding to one soliton solution. 
The initial condition is  $z(t=0) = 0.0239 i $ and $\dot{z}(t=0)=  -43.0768 - 0.00103378 i$
%$\dot{z}(t=0) = -43.0542 - 0.00157258 i$ 
(which is fixed by Eq. \ref{zjdot}) and the equation it satifies is $\ddot{z} = -V^{\prime}(z)$ where %$V(z) =z^2 + 0.12 z^3 + 0.0036 z^4$. 
$V(z) = 0.93 z +0.5z^2 + 0.06 z^3 + 0.0018 z^4$ again using first row of Table 3. 
\textbf{(Upper Panel Right)} Two soliton solution where initial conditions are dictated by two values of dual variable, $z_1 = +  0.0239 i$ and  $z_2 = -  0.0239 i$. We use the prepotential $w(x) = x +0.06 x^2$ which makes the potential
 %$V(x) =30x+x^2 + 0.12 x^3 + 0.0036 x^4$. 
$V(x)= 0.84 x + 0.5 x^2 + 0.06 x^3 + 0.0018 x^4$ using first row of Table 3. We see clear evidence of two solitons in the system. 
\textbf{(Lower Panel Right)} Motion of two coupled dual variables $z_1$ (black) and $z_2$ (blue) is shown here. They satisfy the equation $\ddot{z}_1 = g^2 \frac{1}{(z_1-z_2)^3} -V^{\prime}(z_1)$ where  %$V(z_1) =z_1+z_1^2 + 0.12 z_1^3 + 0.0036 z_1^4$ 
$V(z) =0.9 z_1+0.5 z_1^2 + 0.06 z_1^3 + 0.0018 z_1^4$ 
(similarly for the other dual variable $z_2$). The initial conditions are  $z_1(t=0) = + 0.0239 i $, $z_2(t=0) = - 0.0239 i $, $\dot{z}_1(t=0) = -64.297 - 0.00129161 i $, $\dot{z}_2(t=0) = 64.297 - 0.00129161 i$. 
}
\label{scalingplot}
\end{figure}

\section{Hydrodynamic Limit and Meromorphic Fields}
 \la{sec:mero}
%%%%%%%%%%%%%%%%%%%%%%%%%%%%%%%%%%%%%%%%

\subsection{General formalism}

In this section we consider the generalized Calogero models with external potentials and take the hydrodynamic limit to derive soliton solutions for the corresponding fluid mechanical density and velocity of the particles. We do this by introducing specific meromorphic functions with poles on the position of particles and solitons and taking their many-particle limit. The approach is related to the one of \cite{2005-AbanovWiegmann,2009-AbanovBettelheimWiegmann},
but we will give an independent simplified exposition, directly following from our first-order formulation.

We consider a system with $N$ (real) particle coordinates $x_j$ and $M$ (complex) solitons $z_\alpha$. We will
take the prepotential to be of the form that ensures a stable potential and absence of 3-body forces, as found in
section 2, leading to the first-order equations (\ref{xadot}) 
\be
{\dot x}_a = i \sum_b m_b {\tilde f}_{ab} + i {\tilde w}_a
\label{xdagain}\ee
and corresponding second-order equations
\be
\hspace*{-0.28cm}
{\ddot x}_a = -{1\over m_a} \partial_a V = -\partial_a \left[ \sum_{b (\neq a)} \half m_b (m_a + m_b) {\tilde f}_{ab}^2
+ (m_{tot} - m_a ) v_a + \half {\tilde w}_a^2 \right]
\label{xdd}
\ee
with $v_a = v (x_a )$ and $ {\tilde w}_a= {\tilde w}(x_a)$ as found in section 2.2. 
Coordinates $x_a$ run over particles (for $a = j$) and solitons ($a=\alpha$), and we take the masses of particles to be
$m_j =1$ and the masses of solitons $m_\alpha = -1$, so $m_{tot} = N-M$.

To this system we add one more ``spectator" particle $a=0$ with coordinate $x_0 =x$ and mass $m_0$. The full system
of $N+M+1$ particles and total mass $N-M+m_0$ retains its Calogero-like form. 
Using Eq. \ref{xdagain} for this spectator particle, the velocity $u$ of the spectator particle is, in particular,
\bea
u = {\dot x} &=& i\sum_{a \neq 0} m_a {\tilde f} (x - x_a ) + i {\tilde w} (x)\nonumber \\
&=& i\sum_{j=1}^N {\tilde f} (x - x_j ) - i \sum_{\alpha=1}^M {\tilde f} (x - z_\alpha ) + i {\tilde w} (x)
\label{u(x)}
\eea
and its corresponding acceleration (using Eq. \ref{xdd}) is
\bea
{du \over dt} = {\ddot x} &=& -\partial_x \left[ \sum_{j=1}^N {1+m_0 \over 2} {\tilde f} (x - x_j )^2
+ \sum_{\alpha=1}^M {1 -m_0 \over 2} {\tilde f} (x - z_\alpha )^2 \right. \nonumber\\
&& ~~~~~~~~~+ \left. (N-M)\, v (x) + \half {\tilde w} (x)^2 \right]
\eea

The additional particle $x$ creates an additional term $m_0 {\tilde f} (x_a - x )$ in the equation for 
${\dot x}_a$ of the remaining particles, and a corresponding term in the potential. We wish this particle to be a spectator, that is, not to modify
the motion of the remaining particles (while itself being influenced by them). So we take the limit $m_0 \to 0$, which leaves the
$N$ particles and solitons the same as in the original $N+M$-particle Calogero-like system. Equation (\ref{u(x)}) for $u = {\dot x}$
remains unchanged, while the equation for ${\ddot x}$ becomes
\bea
{du \over dt} &=& -\partial_x \Bigg[ \sum_{j} \half {\tilde f} (x - x_j )^2
\nonumber \\ &+& \sum_{\alpha} \half {\tilde f} (x - z_\alpha )^2 + (N-M) v (x) + \half {\tilde w} (x)^2 \Bigg]
\label{udd}
\eea

The role of the spectator particle is that it monitors and essentially determines both the position and the velocity of the remaining
particles. To this end, we consider $u$ as defined in (\ref{u(x)}) as a function of the spectator particle position $x$ and promote
$x$ to an independent variable, defining a field $u(x)$. The time derivative of $u(x)$, written as ${\partial u \over \partial t}$, is thus
the time variation of $u$ arising from its dependence on $x_j (t)$ and $z_\alpha (t)$, but {\it not} on $x$.
In other words, we define, 
\bea
 {\partial u \over \partial t}  \equiv \sum_j {\partial u \over \partial x_j} {\dot x}_j
+ \sum_\alpha {\partial u \over \partial z_\alpha} {\dot z}_\alpha
\label{ttdef}
\eea

The total time derivative entering (\ref{udd}) is, therefore,
\bea
{du \over dt} &=& {\partial u \over \partial x} {\dot x} + \sum_j {\partial u \over \partial x_j} {\dot x}_j
+ \sum_\alpha {\partial u \over \partial z_\alpha} {\dot z}_\alpha \nonumber \\
&=& u \partial_x u + {\partial u \over \partial t} = \partial_x \left(\half u^2 \right) +  {\partial u \over \partial t} 
\label{ttd}
\eea
where we used ${\dot x} = u$ and the definition Eq. \ref{ttdef}.

The above relation and (\ref{udd}) allow us to find the equation of motion of
the field $u(x,t)$. To do this, we need to express the terms involving $x_j$ and $z_\alpha$ in (\ref{udd}) in terms of
$u$. This can be achieved by noting that all prepotentials ${\tilde f} (x)$ found in section 2.2 
(rational, trigonometric or hyperbolic) satisfy the relation
\be
{\tilde f}(x)^2 = g \, \partial_x {\tilde f} (x) + C ~,~~~ f(x) = -{g \over x}, ~ -g\cot x , ~ -g \coth x
\label{der}
\ee
where $C$ is a constant 
(zero, $+g^2$ and  $-g^2$ for rational, hyperbolic and trigonometric prepotential respectively). Therefore
the terms involving sums of ${\tilde f}^2$ in (\ref{udd}) can be expressed as derivatives of terms
in $u(x)$. We note, however, that particle and soliton terms come with opposite sign in $u(x)$, while they
have the same sign in (\ref{udd}). This necessitates splitting $u(x)$ into two parts:
\bea
u^+ (x) &=& - i \sum_{\alpha} {\tilde f} (x - z_\alpha ) + i \lambda {\tilde w} (x)  \label{up}\\
u^- (x) &=& i\sum_{j} {\tilde f} (x - x_j ) + i (1-\lambda) {\tilde w} (x) \label{um}\\
u(x) &=& u^+ (x) + u^- (x) \nonumber
\eea
where $\lambda$ is an arbitrary parameter that splits the term ${\tilde w} (x)$ between the two
functions. Using (\ref{udd},\ref{ttd}, \ref{der}) and (\ref{up},\ref{um}) we arrive at the equation of motion for $u$
\bea
{\partial u \over \partial t} &+& \partial_x \Bigg[ \half u^2 + {i g \over 2} \partial_x ( u^+ - u^- ) \nonumber \\ &+& \half {\tilde w}^2 
+ (N-M) v +(\lambda - {\textstyle \half}) \, g \,\partial_x {\tilde w}\Bigg] = 0
\eea
The terms independent of $u$ above are the one-body potential $V(x)$ entering the equation of motion of
particles $x_j$, with the difference that $v(x)$ is multiplied by $N-M$ (rather than $N-M-m_j = N-M-1$
and the extra term
involving $\partial_x {\tilde w} (x)$. In fact, for all cases of $\tilde f$ (rational, trigonometric and hyperbolic),
$\partial_x {\tilde w} (x)$ is proportional to $v(x)$:
\be
g \, \partial_x {\tilde w} (x) = -2 v(x)
\ee
The equation for $u(x,t)$ therefore takes the form:
\be
{\partial u \over \partial t} + \partial_x \left[ \half u^2 + {i g \over 2} \partial_x ( u^+ - u^- ) +V
+ (\lambda -1)g \, \partial_x {\tilde w} \right] = 0
\label{ut}\ee

If we want the equation of motion of $u$ to involve the same one-body potential $V(x)$ as that of
particles, we must choose $\lambda =1$, assigning the full prepotential ${\tilde w} (x)$ to $u^+$.
This is in fact the opposite convention than the one in \cite{2009-AbanovBettelheimWiegmann}, where
$u^-$ contains the term  ${\tilde w} (x)$. We stress that any choice of $\lambda$ is allowed,  leading to
different definitions of $u^+$, $u^-$ and an extra term in the evolution equation for $u$.
We also note that $\lambda$ does not appear in the equation of $u$ (\ref{ut}) in the quadratic (harmonic)
rational Calogero case studied in
Ref. \cite{2009-AbanovBettelheimWiegmann}, since $w(x)$ is
linear and $v(x)$ is a constant that drops from the equation.

\subsection{The rational case and derivation of the hydrodynamic limit}

The above construction holds for all three types of Calogero potentials. We first examine the rational case.
We pick $\lambda=1$ as the most natural choice and define
\bea
u^+ (x) &=& ig\sum_{\alpha=1}^{M}\frac{1}{x-z_{\alpha}} + i {\tilde w} (x) \label{upp}\\
u^- (x) &=& -ig\sum_{j=1}^{N}\frac{1}{x-x_j}\label{umm}
\eea
A priori, it looks like we have a single equation of motion (\ref{ut}) for two fields $u^+$ and $u^-$.
As we stressed before, however, the particle system is actually fully determined by the values of $z_\alpha$,
so in principle $u^+$ is enough to fully fix the system.
$u^+$ is a meromorphic function of $x$ with $M$
simple poles at $z_\alpha$, so it fixes the number of solitons. 
Using the equations of motion (\ref{xjdot}) we see that $u^+$ satisfies
\be
    u^+(x_j) = {\dot x}_j + ig\sum_{k=1 (k\neq j)}^{N} \frac{1}{x_j-x_k}.
    \la{u+j}
\ee
Therefore, {\it if we also know that there are $N$ particles}, the function $u^+ (x)$ fully determines the system, as:
\bea
   && Im ~u^{+}(x_j ) = g\sum_{k=1 (k\neq j)}^{N} \frac{1}{x_j-x_k} \label{Iu}
  \\
&& Re ~ u^{+}(x_j ) = {\dot x}_j = v_j
\la{Ru}
\eea
The $N$ equations (\ref{Iu}) in principle determine the $N$ real variables $x_j$, and subsequently
(\ref{Ru}) determines ${\dot x}_j$. The known values of $x_j$, then, determine the function $u^- (x)$ through
equation (\ref{umm}).

The definition  of $u^+ (x)$, (see Eq. \ref{upp}) however, does not involve $N$, and the {\it same} function $u^+ (x)$
can describe systems of an arbitrary number of particles (see Eq. \ref{Iu}, \ref{Ru}). The real part of $u^+ (x)$ for real $x$, in particular, 
defines a continuous velocity field
$v(x)$ that is the actual particle velocity on the position of particles. Similarly, its imaginary part defines
a field that can be related to the position of particles, for any number of them. It is, therefore, a good tool to
deal with the hydrodynamic limit $N \to \infty$ where the interparticle distance goes to zero and the system is
described by a continuous density $\rho (x)$ and velocity $v(x)$.

From the above discussion it follows that, in the $N \to \infty$ limit, the real part of $u^+ (x)$ straightforwardly
goes over to the fluid velocity field $v(x)$. To express the imaginary part in terms of the fluid particle density
requires a bit more work. In particular, we need to express the sum in (\ref{Iu}) in terms of the particle
density $\rho (x)$, including all perturbative corrections in $1/N$. This is nontrivial because of the singularity
as $x_k$ approaches $x_j$ and has to be evaluated carefully. This has been done in \cite{Stone}. Here we will
follow a slightly different approach that will allow us to separate the perturbative and non-perturbative parts.

For a large number of particles we define the continuous position function $x(s)$ such that $x(j) = x_j$.
For finite $N$ any smooth interpolation between the $x_j$ will do, while in the
$N \to \infty$ limit this function becomes unique. It is related to the continuous density $\rho (x)$ by
\be
x' (s) = {1 \over \rho (x(s))}
\ee
The sum of interest is
\be
\sum_{k(\neq j )=1}^N {1 \over x_j - x_k} = \sum_{k(\neq j)=1}^N {1 \over x(j) - x(k)}
\la{sum}
\ee
which needs to be expressed in terms of $x(s)$ or $\rho(x)$.

Our starting point is the identity
\be
\sum_{n=-\infty}^\infty f(n) = \sum_{n=-\infty}^\infty {\tilde f} (2\pi n)
\ee
where $f(s)$ is any function of $s$ and ${\tilde f} (q)$ is its Fourier transform, defined as
\be
 {\tilde f} (q) = \int_{-\infty}^\infty ds~ e^{-i q s} f(s)
\ee
For a function smooth at the scale of $\Delta s \sim 1$, the Fourier transform ${\tilde f} (2\pi n)$ for
$n \neq 0$ will be negligibly small. In fact, terms with $n\neq 0$ are nonperturbative in $1/N$
(instanton corrections), as we will explain in the sequel. So, up to nonperturbative corrections
\be
\sum_{n=-\infty}^\infty f(n) =  {\tilde f} (0) =  \int_{-\infty}^\infty ds ~f(s)
\la{sumint}
\ee
To apply this formula to the sum (\ref{sum}) we view the summand as a function of $k$ and define it to be zero for $k$
outside its range, extending the summation to all integers. Still, a straightforward application of (\ref{sumint}) 
is hindered by
the fact that the summed function $1/[x(j) - x(s)]$ is not smooth, due to the singular behavior near $s=j$, and thus
higher Fourier modes contribute substantially. We can proceed in two different ways. The first is to evaluate
the higher Fourier modes and sum their contribution. The second is to regularize the integrand in a way that
renders it smooth, as was done in \cite{Stone}. Both methods lead to the
same result. Following the second method, we define the function
\be
f(s) = {1 \over x(j) - x(s)} + {1 \over x' (j) (s-j) + \epsilon (s-j)^3} ~,~~ f(j) = {x'' (j) \over 2 x' (j)^2}
\ee
for $\epsilon >0$. The above function is continuous everywhere and smooth at the scale of $\Delta s \sim 1$,
since the
singularity at $s=j$ is subtracted by the second (regulator) term. Summing it over integer values of $s$ we have
\bea
\sum_{k=-\infty}^\infty f(k) &=& f(j) + \sum_{k(\neq j)=-\infty}^\infty f(k) \nonumber\\
&=&{x'' (j) \over 2 x' (j)^2} + \sum_{k(\neq j)=-\infty}^\infty {1 \over x(j) - x(k)}
\eea
since the sum of the regulator term is absolutely convergent (due to $\epsilon$) and
vanishes due to antisymmetry in $k-j$. Applying (\ref{sumint}) we have the result
\bea
\sum_{k(\neq j)=-\infty}^\infty {1 \over x(j) - x(k)} &=& -{x'' (j) \over 2 x' (j)^2} + \int_{-\infty}^\infty
f(s) ds \nonumber \\
&=&  -{x'' (j) \over 2 x' (j)^2} + P.V. \int_{-\infty}^\infty {ds \over x(j) - x(s)}
\eea
where $P.V.$ stands for principal value. Finally, changing integration variable from $s$ to $y= x(s)$
and using $x' = 1/\rho$ and thus $x'' = -\partial_x \rho / \rho^3$ we obtain
\bea
\sum_{k(\neq j)=-\infty}^\infty {1 \over x - x(k)}  &=& {\partial_x \rho (x) \over 2 \rho (x)}
+ P.V. \int_{-\infty}^\infty {\rho (y) dy \over x - y} \nonumber \\
&=& \half \partial_x \ln { \rho(x)} - \pi \rho^H (x)
\eea
where we put $x(j) = x$, and $\rho^H (x)$ is the Hilbert transform of $\rho(x)$.

The above result is perturbatively exact in $1/N$. Indeed, the higher Fourier modes of $f(s)$ expressed
in terms of $y = x(s)$ are
\be
 {\tilde f} (q) = \int_{-\infty}^\infty dy \, \rho(y) ~ e^{-i q \int_{-\infty}^y dw \, \rho(w)} \, f(s(y))
\ee
For a continuous particle distribution, $x(j+1) - x(j) \sim 1/N$, $\rho(x) \sim N$, and thus the
oscillating exponent in the above expression for $q \neq 0$ is of order $N$. The integral is thus of order
$e^{-N}$, which is nonperturbative in $1/N$. So the $q=0$ term captures the full perturbative contribution.
Overall, from (\ref{Iu},\ref{Ru}) we obtain for $u^+ (x)$
\be
 \la{u+td}
	u^+ (x)=v(x)-i\pi g \rho^H (x) +ig\, \partial_x\ln {\sqrt{\rho (x)}}
\label{u+rv}
\ee

Nonperturbative contributions are in general negligible in the fluid limit ($N \to \infty$). The one instance
in which they become relevant is when the distribution of particles breaks into two or more disjoint components.
In this case the distance $x(K) - x(K +1)$ between the last particle $K$ in one component and the first particle 
$K+1$ in the next is large, and thus the function $f(s)$ is not smooth at $s=K$.
The appearance of multiple fluid components signals the onset of nonperturbative effects and needs to
be described in terms of multiple functions $\rho_a (x)$, one for each component, with compact disjoint supports. In our paper, we do not encounter this scenario and hence, emergence of relavant perturbative corrections is not a concern. 

The continuous version of $u^- (x)$ can similarly be found. Its definition (\ref{umm}) involves the parameter
$x$ and a sum over the full set of particles. By taking the variable $x$ to be complex and off the real axis the
issue of singularities is avoided and the summand becomes a smooth function of $x(s)$. So, up to nonperturbative
contributions,
\be
u^- (x) = -ig\sum_j \frac{1}{x-x(j)} = -ig \int ds \, \frac{1}{x-x(s)} = -ig \int dy\, {\rho(y) \over x- y}
\ee
So $u^- (x)$ is the Cauchy transform of $\rho(x)$. As $x$ approaches the real axis the above expression
has a discontinuity. We obtain
\be
 \la{u-td}
	u^-(x\pm i0)=\mp \pi g \rho + i\pi g \rho^H
\label{u-r}
\ee
with the discontinuity
\be
	u^{-}(x+i0)-u^{-}(x-i0)=-2\pi g \rho(x).
 \la{u-jump}
\ee

Expressions (\ref{u+rv}) and (\ref{u-r}) determine $u^\pm (x)$ in terms of fluid  quantities. Note that with our choice of $\lambda=1$ in the definition (\ref{up},\ref{um}) of $u^\pm (x)$
their expression involves only the fluid density $\rho(x)$ and velocity $v(x)$ and not the prepotential ${\tilde w} (x)$.
Substituting these expressions into the equation (\ref{ut}) for $u(x)$ we obtain in principle 4 real equations 
(2 for the real part and 2 for the imaginary part at $x\pm i0$) for the two real fields $\rho(x)$ and $v(x)$.
These equations are compatible and reduce to the fluid equations
\bea
\partial_t \rho &+& \partial_x (\rho v ) = 0 ~~~~~~~~~~~~~~~~~~~~~~~~~~~~~~~~~~~\\
\partial_t v &+& \partial_x \Bigg[ \half v^2  + {\pi^2 g^2 \over 2} \rho^2 + \pi g^2 \partial_x \rho^H
\\ \nonumber &-& {g^2 \over 8} \left( \partial_x \ln { \rho}\right)^2 - {g^2 \over 4}
\partial_x^2 \ln {\rho}  + V \Bigg]= 0
\eea
The above equations can be seen to arise from the Hamiltonian
\be
H = \int dx \left[ \half \rho v^2 +{\pi^2 g^2\over 6} \rho^3 +{\pi g^2\over 2} \rho \, \partial_x \rho^H
+ {g^2} {(\partial_x \rho)^2 \over 8\rho} + \rho V \right]
\ee
using the standard fluid mechanical Poisson structure
\be
\{ \rho (x) , v (y) \} = \delta' (x-y)
\ee

\subsection{One soliton solution of Calogero model with quartic potentials in terms of meromorphic fields}

The one-soliton solution is given by
\be
	u^{+}(x) = \frac{ig}{x-z_{1}(t)} + i {\tilde w} (x),
 \la{u+1sol}
\ee
The single soliton solution for $u^+ (x)$ is a meromorphic function with a single pole. $z_{1}(t)$ above satisfies (\ref{zjdot}) for $M=1$ which is simply
\be
\dot{z}_1 -i{\tilde w} (z_1 ) = -ig\sum_{j=1}^{N}\frac{1}{z_1-x_{j}}
\ee
and the second-order equation
\be
{\ddot z}_1 = - V' (z_1 )
\label{onezquart}
\ee
From the expression of $u^+ (x)$ in terms of hydrodynamic quantities
\be
 \la{TDred1}
	v - ig(\pi\rho^H-\partial_x\log\sqrt{\rho})=\frac{ ig}{x-z_{1}} + i {\tilde w} (x) 
\ee
we can, in principle, find the density and velocity fields from the position $z_{1}$. Writing $z_1 (t) = a(t) + i b(t)$
and taking the real and imaginary parts of (\ref{TDred1}) we obtain 
\bea
	v&=&-g \, \frac{b}{(x-a)^2 + b^2} \la{1solTDv}\\
	\pi\rho^H &-&  \partial_x\log\sqrt{\rho}
	= -\frac{x-a}{(x-a)^2 + b^2} - {1 \over g} {\tilde w} (x)
 \la{1solTD}
\eea
So $a(t)$ parametrizes the position of the soliton while $b(t)$ parametrizes its width.
The second equation above needs to be solved for $\rho (x)$. This is nontrivial,
although the solution can be found analytically in the limit of thin solitons. In this limit, the width $b \rightarrow 0$, and the soliton solution (upto $O(1/N)$ corrections), can be written as 
\bea
\rho_{sol} (x,t)= \rho_{0} + \delta(x - a(t))
\eea
where the background density $\rho_{0} $ satisfies, $\pi\rho_0^H -  \partial_x\log\sqrt{\rho_0}
	+ {1 \over g} {\tilde w} (x)=0
$

The soliton parameter $z_{1}(t)$ is moving in the complex plane along a non-trivial curve guided by its quartic polynomial
as in (\ref{onezquart}). Therefore, (\ref{1solTDv},\ref{1solTD}) give a one-dimensional reduction of an infinite dimensional rational Calogero system with quartic potential in the hydrodynamic limit. The procedure to go to the hydrodynamic limit can similarly be extended for the two-soliton and multi-soliton case. 
%\vskip 0.1in
\vskip 0.1in
\section{Conclusions and Outlook}
 \la{sec:conclusion}

%To summarize, in this paper, we introduced a first order formalism and derived the generalized form of two-body and external potentials. Imposing the requirements of stability and reality conditions, we demonstrated the natural emergence of Calogero family of models in generalized quartic potentials. Our general formalism provides a relatively easy route to finding solitons which is otherwise considered to be an enormous challenge. Using the more common version of the Calogero family of models, namely, the rational Calogero model (in quartic polynomial external potentials), we demonstrate the existence of solitons. We showed the particle time evolution for the case when the system has one and two solitions and we showed that our method can be easily extended to M-solitons. We showed that finding soliton solutions can be achieved via a mapping to an electrostatic problem. Using a collective fluid formalism and the notion of meromorphic fields, we have also found soliton solutions in the hydrodynamic limit. 

To summarize, in this paper we introduced a first order formalism based on a prepotential and derived its general form that gives rise to two-body and external potentials. Imposing the requirements of stability and reality conditions,
we demonstrated the natural emergence of the Calogero family of models in generalized quartic and trigonometric external potentials. Our general formalism provides a relatively straightforward route to finding soliton solutions,
a task otherwise considered to be an enormous challenge. Using the more common version of the Calogero family of models, namely, the rational Calogero model (in quartic polynomial external potentials), we demonstrate the existence of soliton solutions. We derived the particle time evolution for the case when the system has one and two solitons and we showed that our method can be easily extended to $M$ solitons. We showed that finding soliton solutions can be achieved via a mapping to an electrostatic problem. Using a fluid formalism involving meromorphic fields, we have also identified soliton solutions in the hydrodynamic limit.

One of the the main lesson from the work presented in this paper is that there may exist further extensions of the Calogero
family of models beyond the known systems, and that they may admit dual formulations that identify their collective
degrees of freedom and provide solutions to their fluid mechanical versions. Clearly, there are many open issues and
directions of possible future research.

The most immediate questions are the ones on stability and integrability.
It is puzzling that the dual formulation of the quartic potential model is stable only within a subset of its parameters, which
actually exclude the purely quartic case. Although we conjecture that models outside the stability regime correspond to a
soliton condensate, an explicit demostration of this fact, and derivation of the soliton solutions, would be desirable.

Similarly, our approach does not deal with integrability. Again, it is remarkable that the systems that can be
dealt with this formalism do fall eventually within a subclass of the generalized Calogero models that were known to 
be integrable.
A direct derivation of integrability seems to be possible within this formalism and, if there, has yet to be uncovered.

Extension of our results to other members of the Calogero family is also an open issue. We restricted our derivation to
the rational, trigonometric and elliptic models and their external potential generalizations, mainly for reasons of
mathematical clarity and simplicity. An extension to the elliptic (Weierstrass) model is well within the reach of the
formalism. In this context, the identification of elliptic models with external potentials would be a very interesting
advance. Similar remarks hold for models of particles with internal degrees
of freedom. Clearly an extension of the formalism is needed to incorporate internal particle coordinates, and this is a topic
of further research.

Finally, there exist several intriguing similarities of the present formalism with quantum mechanical features of the
Calogero model, although our treatment is purely classical. The generating function clearly alludes to a quantum
mechanical wavefunction, at least in the equilibrium semiclassical limit. Similarly, the stable and unstable domains of
quartic dual systems are in direct analogy with the broken and unbroken phases of supersymmetric quantum mechanical systems,
the ``unstable" broken phase leading to a soliton condensate. Aspects of our formalism also bear similarities with
techniques from matrix models and the exchange operator formulation. These and related issues are left for future investigation.

\section{Acknowledgments}
%%%%%%%%%%%%%%%%%%%%%%%%%%%%%%%%%%%%%%%%%%%%%%%%%%%%%%%%%%%%
M. K. gratefully acknowledges the Ramanujan Fellowship SB/S2/RJN-114/2016 from the Science and Engineering Research Board (SERB), Department of Science and Technology, Government of India. A.P.'s research is supported by NSF grant 1213380 and by a PSC-CUNY grant.

%%%%%%%%%%%%%%%MANAS COMMENTED BELOW BELOW BELOW %%%%%%%%%%%%%%%%%%%%%%%%%%
%%%%%%%%%%%%%%%%%%%%%%%%%%

%%%%%%%%%%%%%%%%%%%%%%%%%%%%%%%%%%%%%%%%%
%%%%%%%%%%%%%%%%%%%%%%%%%%%%%%%%%%%%%%%%%
\section{Appendix A: Solutions to the functional equations without external potentials}
\la{app-hilbert}
In this Appendix we provide the solutions to the functional equations for ${\tilde f}_{ab}$, thereby yielding Table 1. 
We start with the equation (\ref{functf}) for a fixed triplet of indices $b,c,d$:
\bea
\tilde{f}_{bc}\tilde{f}_{bd}
-\tilde{f}_{bc}\tilde{f}_{cd}+\tilde{f}_{bd}\tilde{f}_{cd}=C_{bcd}
\eea
where we used the antisymmetry of ${\tilde f}_{ab} = - {\tilde f}_{ba}$ to put the indices $b,c,d$ in order.

Let us define, $x\equiv x_b - x_c$,  $y\equiv x_c - x_d$ such that $x+y = x_b-x_d$.
For convenience, we also rename the doublets of indices $bc \equiv 1$, $cd \equiv 2$, $ bd \equiv 3$ and call
$C_{bcd}$ simply $C$. The above equation, then, reads
\bea
\tilde{f}_{1} (x)\tilde{f}_{3}(x+y) - \tilde{f}_{1}(x) \tilde{f}_{2}(y)+\tilde{f}_{3}(x+y) \tilde{f}_{2}(y)=C
\eea
In terms of the reciprocal functions $g_j (x)=\frac{1}{{\tilde f}_j (x)}$ the above becomes 
\bea
g_1 (x) + g_2 (y)- g_3 (x+y) = C g_1 (x) \, g_2 (y) \, g_3 (x+y)
\eea
and solving for $g_3$ we obtain
\bea
g_3 (x+y) = \frac{g_1 (x) +  g_2 (y)}{1 + C g_1 (x) g_2 (y)}
\label{g3}
\eea
Taking derivatives with respect to $x$ and $y$ and equating the results 
(since $\partial_x g_3 (x+y) = \partial_y g_3 (x+y)$) we obtain
\bea
\frac{g_1^{\prime} (x) }{1-C g_1^2 (x)} = \frac{g_2^{\prime} (y)}{1-C g_2^2 (y)} = k = 
{\rm constant}
\label{g12}
\eea
The above differential equations determine $g_1$ and $g_2$ and, through (\ref{g3}),
they also determine $g_3$. Their solution depends on the sign of the constant $C$. Specifically
\bea
C=0:&  &g_j (x) = k x + b_j \nonumber\\
C= -g^2 < 0 :&  &g_j (x) = \frac{1}{g} \sin (kg x +g \, b_j )  \nonumber \\
C= g^2 > 0 :&   &g_j (x) = \frac{1}{g} \sinh (kg x + g \, b_j ) \nonumber \\
j=1,2,3,& &{\rm with}~~     b_1 + b_2 = b_3
\eea
Repeating the above analysis for a triplet involving one new particle, say, $b,c,e$, leads to the
same equations (\ref{g12}), with the same constants $C$ and $k$ (fixed by the form of $g_{bc} = g_1$
found above). Inductively, we conclude that the constants $C=C_{abc}$ and $k$ are common
for the full set of particles, while the constants $b_{ab}$ satisfy $b_{ab} + b_{bc} = b_{ac}$
for all $a,b,c$. These constants can actually be absorbed by a shift in the position of particle
coordinates:
\be
x_a \to x_a - \frac{b_{1a}}{k} ~,~~ b_{11} \equiv 0
\ee
where we chose arbitrarily particle $1$ as a reference, so we can take all $b_a =0$.
Finally, the constant $k$ can be set to $1/g$ through the rescaling of coordinates 
$x_a \to (kg)^{-1} x_a$. Overall, we recover the ${\tilde f}_{ab} = 1/g_{ab}$ as given
in Table 1.
%%%%%%%%%%%%%%%%%%%%%%%%%%%%%%%%%%%%%%%%

\section{Appendix B: Solutions to functional equations for external potentials}

In this Appendix we derive the solutions to the functional equation (\ref{bc}), written explicitly as
\be
 \big[{\tilde w}_b (x_b ) - {\tilde w}_c (x_c )\big] \,{\tilde f}_{bc} (x_{bc} )= u_{bc} (x_{bc} ) + v_b (x_b )+ v_c (x_c )
\ee
We define $x_b = t+s$ and $x_c = t-s$, which implies $x_{bc} = 2s$. Using also the inverse functions
$g_{bc} = 1/{\tilde f}_{bc}$ defined in the previous Appendix, the above equation becomes
\bea
{\tilde w}_b (t+s) - {\tilde w}_c (t-s) &=&  g_{bc} (2s) \bigg[ u_{bc} (2s)+ v_b (t+s)+ v_c(t-s) \bigg]\nonumber \\
&=& h_{bc} (s) +  g_{bc} (2s) \bigg[v_b (t+s)+ v_c(t-s) \bigg]
\label{bcapp}
\eea
where we defined $h_{bc} (s) = g_{bc}  (2s) u_{bc} (2s)$.
The solution of this equation depends on the form of $g_{bc}$ and we treat it in a case-by-case basis for
the solutions derived in the previous Appendix.  

\subsection{The rational case  $g_{bc}={1 \over {\tilde f}_{bc} } = \frac{1}{g} x_{bc}$}

In the rational case (\ref{bcapp}) becomes 
\bea
{\tilde w}_b (t+s) - {\tilde w}_c (t-s) &=& {2s \over g} \bigg[ u_{bc} (2s)+ v_b (t+s)+ v_c(t-s) \bigg]
\nonumber \\
&=& h_{bc} (s) +  {2s \over g} \bigg[v_b (t+s)+ v_c(t-s) \bigg]
\label{wrat}
\eea
with $h_{bc} (s) = 2s \, u_{bc} (2s)/g$.
The left hand side is regular around $s=0$ and has a well-defined Taylor expansion in $s$, therefore so must be 
$h_{bc} (s)$. Expanding in powers of $s$ we obtain
\bea
s^0 :&   &{\tilde w}_b (t) -  {\tilde w}_c (t) = h_{bc}(0) \label{Taylor.0} \\
s^1 :&   &{\tilde w}_b' (t) +  {\tilde w}_c' (t) = h_{bc}'(0) +  {2\over g} \big[ v_b (t) + v_c(t) \big]\label{Taylor.1} \\
s^2 :&   &{\tilde w}_b'' (t) -  {\tilde w}_c'' (t) = h_{bc}''(0) +  {4\over g} \big[ v_b' (t) - v_c' (t) \big]\label{Taylor.2}\\
s^3 :&   &{\tilde w}_b''' (t) +  {\tilde w}_c''' (t) = h_{bc}'''(0) +  {6\over g} \big[ v_b'' (t) + v_c'' (t) \big] \label{Taylor.3}
\eea
Eq.\ (\ref{Taylor.0}) states that ${\tilde w}_b (t)$ and ${\tilde w}_c (t)$ differ by a constant. 
Differentiating (\ref{Taylor.1}) twice with respect to $t$ and combining with (\ref{Taylor.3}) we obtain
\be
 {\tilde w}_b''' (t) = -{1\over 4} h_{bc}''' (0) = {\rm constant}
\ee
which means that ${\tilde w}_{b}$ and ${\tilde w}_{c}$ must be of the form
\be
{\tilde w}_{b,c} (t) = C_{b,c} + c_1 t + c_2 t^2  + c_3 t^3
\ee
The constants $C_b$ and $C_c$ can differ, but $c_1 , c_2 , c_3$ are the same for any two particles $b$ and $c$,
therefore they are common to the system. Substituting this form for ${\tilde w}_{b,c} (t)$ in (\ref{Taylor.0}-\ref{Taylor.3}),
or directly in (\ref{wrat}), we obtain $v_b (x)$, $v_c (x)$, $h_{bc}(x)$ and $u_{bc} (x)$. In doing this, we note  that
the constant terms of $v_b$, $v_c$ and $u_{bc}$ can be combined together; similarly, $v_b' (0) - v_c' (0)$ and
$u_{bc}' (0)$, contributing a term proportional to $x_b - x_c = x_{bc}$, can also be combined. We use this to choose
$v_b (0) = v_c (0) = v_b' (0) - v_c' (0) = 0$. We eventually obtain
\bea
{\tilde w}_{b,c} (x) &=& C_{b,c} + c_1 x + c_2 x^2  + c_3 x^3 \\
v_{b,c} (x) &=& g c_2 x + {3g \over 2} c_3 x^2 \\
u_{bc} (x) &=& g  {C_b - C_c \over x_{bc}} + g c_1 -{g \over 2} c_3 x^2
\eea
The above recover the potentials presented in Table 2. Note that the constant $g c_1$ in $u_{bc}$ is
dynamically irrelevant and can be omitted. Note also that the contribution  of $v_a (x_a )$ in the full
potential involves a sum $\sum_{a\neq b} m_a m_b v_a = \sum_a (m_{tot} - m_a ) m_a v_a$, which explains the
coefficient of the corresponding terms in $V_a (x_a )$. 

\subsection{The trignometric case  $g_{bc} = {1 \over {\tilde f}_{bc} } = {1\over g} \tan x_{bc}$}

In the trignometric or hyperbolic case we proceed in a similar way, Taylor expanding (\ref{bcapp}) in $s$. The equations
are the same as
in the previous section for orders $s^0$, $s^1$ and $s^2$, since $\tan x$ is the same as
$x$ up to quadratic order. At order $s^3$, however, we get instead of (\ref{Taylor.3})
\be
{\tilde w}_b''' (t) +  {\tilde w}_c''' (t) = h_{ab}'''(0) +  {6\over g} \big[ v_b'' (t) + v_c'' (t) \big] 
+{16 \over g} \big[ v_b (t) + v_c (t) \big]
\ee
Combining this with the other equations yields 
 \bea
{\tilde w}_b''' (t) + 4 {\tilde w}_b' (t) &=& -{1 \over 4} h_{bc}''' (0) + 2 h_{bc}' (0) \equiv {\rm C}\nonumber\\
{\rm or} ~~ {\tilde w}_b'' (t) + 4 {\tilde w}_b (t) &=& {\rm C} t + {\rm C}'
\eea
This is like a driven harmonic oscillator of frequency $2$, with general solution of the form
\be
w_{b,c} (t) = C_{b,c} + c_1 \cos 2t + c_2 \sin 2t + c_3 t
\ee
Putting this form in the remaining equations, or in the original functional equation, we finally find
\bea
{\tilde w}_{b,c} (x) &=& C_{b,c} + c_1 \cos 2x + c_2 \sin 2x  + c_3 x \\
v_{b,c} (x) &=& g \, c_2 \cos 2x - g \, c_1 \sin 2x  \\
u_{bc} (x) &=& g  (C_b - C_c) \cot x + g \, c_3 \, x \cot x
\eea
The hyperbolic case is treated in exactly the same way, or can be obtained by simple analytic continuation
$x \to i x$, $g \to ig$, $c_2 \to -i c_2$. Altogether, we recover the potentials presented in Table 2. 

%%%%MANAS COMMENTED ABOVE ABOVE ABOVE %%%%%%%%%%%%%%%%%%%%%%%%%%%%%%%%%%%%%
\section*{References}
%%%%%%%%%%%%%%%%%%%%%%%%%%%%%%%%%%%%%%%%

%%%%%%%%%%%%%%%%%%%%%%%%


\begin{thebibliography}{99}
%%%%%%%%%%%%%%%%%%%%%%%%%%%%%%%%%%%%%%%
\bibitem{Calogero-1969}
    F. Calogero, J. Math. Phys. \textbf{10}, 2191 (1969).
   \textit{Solution of a Three-Body Problem in One Dimension.}
   ibid.  \textbf{10}, 2197 (1969);
  \textit{Ground State of a One-Dimensional $N$-Body System.}
  ibid. \textbf{12}, 419 (1971).
  \textit{Solution of One-Dimensional $N$-Body Problems with Quadratic and/or Inversely Quadratic Pair Potentials.}

\bibitem{Sutherland-1971}
    B. Sutherland, Phys. Rev. A \textbf{4}, 2019 (1971);
  \textit{Exact Results for a Quantum Many-Body Problem in One Dimension.}
    ibid., \textbf{5}, 1372 (1972).
  \textit{Exact Results for Quantum Many-Body Problem in One Dimension. II.}
    Phys. Rev. Lett. \textbf{34}, 1083 (1975).
  \textit{Exact Ground-State Wave Function for a One-Dimensional Plasma.}

\bibitem{Moser}
J. Moser, Adv. Math. 16, 197 (1975). \textit{Three integrable Hamiltonian systems connected with isospectral deformations}

\bibitem{OlshaPer}
M. A. Olshanetsky and A. M. Perelomov,
Phys. Rep. \textbf{71},  pp. 313-400, (1981),
\textit{Classical integrable finite-dimensional systems related to Lie algebras};
Phys. Rep. \textbf{94}, Issue 6, pp. 313-404 (1983),
\textit{Quantum Integrable Systems Related to Lie Algebras.}


\bibitem{Perelomov-book}
For a review and original references see
A. M. Perelomov, \textit{Integrable Systems of Classical Mechanics and Lie Algebras.}, Birkh\"auser Basel (1989).
%
%\bibitem{PerQuant}
%For a review and original references see
%M. A. Olshanetsky and A. M. Perelomov, Physics Reports, Volume 94, Issue 6, p. 313-404.,
%\textit{Quantum Integrable Systems Related to Lie Algebras.}, Birkh\"auser Basel (1989).

\bibitem{Sutherland-book}
	For a review and original references see B. Sutherland, \textit{Beautiful Models:
	70 Years Of Exactly Solved Quantum Many-Body Problems.},
	World Scientific, (2004).

\bibitem{APreview} Alexios P Polychronakos, Journal of Physics A: Mathematical and General, Volume 39, Number 41, \textit{The physics and mathematics of Calogero particles}


\bibitem{i3}
V.I. Inozemtsev, Physica
Scripta 29 (1984) 518-520.,  \textit{New completely integrable multiparticle dynamical systems}



\bibitem{APnew1}  Alexios Polychronakos, Phys.Lett. B276 (1992) 341-346
\textit{A New integrable system with a quartic potential}

\bibitem{APnew2}  Alexios P. Polychronakos, Phys.Lett. B277 (1992) 102-108
\textit{New integrable systems from unitary matrix models}


\bibitem{Kawakami1} N. Kawakami, Phys. Rev. B46, 1005 (1992). \textit{Asymptotic Bethe-ansatz solution of multicomponent quantum systems with $1/r^2$ long-range interaction}

\bibitem{Kawakami2} N. Kawakami, Phys. Rev. B46,  3191 (1992). \textit{SU(N) generalization of the Gutzwiller-Jastrow wave function and its critical properties in one dimension}

\bibitem{HaHa} Z.N.C. Ha and F.D.M. Haldane, Phys. Rev. B46, 9359 (1992), \textit{Models with inverse-square exchange}

\bibitem{MinahanAP} J.A. Minahan and A. P. Polychronakos, Phys. Lett. B302, 265 (1993) [arXiv:hep-th/9206046]. \textit{Integrable Systems for Particles with Internal Degrees of Freedom}

\bibitem{HiWa} K. Hikami and M. Wadati, Phys. Lett. A173, 263 (1993). \textit{ Integrable spin-12 particle systems with long-range interactions}


\bibitem{APspinchain}
Alexios P. Polychronakos,  Phys.Rev.Lett. 70 (1993) 2329-2331, \textit{Lattice Integrable Systems of Haldane-Shastry Type}

\bibitem{APspinchain1}
Alexios P. Polychronakos, Nucl.Phys. B419 (1994) 553-566, \textit{Exact Spectrum of SU(n) Spin Chain with Inverse-Square Exchange}


\bibitem{JevickiSakita}
      A. Jevicki and B. Sakita, Nucl. Phys. \textbf{B165},
   511 (1980).
   \textit{The Quantum Collective Field Method and its
   Application to the Planar Limit.}

\bibitem{Sakita-book}
    B. Sakita, \textit{Quantum Theory of Many-variable Systems and
Fields.}, World  Scientific, 1985.

\bibitem{Jevicki-1992}
      A. Jevicki, Nucl. Phys. \textbf{B376}, 75-98 (1992).
   \textit{Nonperturbative Collective Field Theory.}

\bibitem{2005-AbanovWiegmann}
    A. G. Abanov and P. B. Wiegmann, Phys. Rev. Lett {\bf 95}, 076402 (2005).
  \textit{Quantum Hydrodynamics, the Quantum Benjamin-Ono equation, and the Calogero Model.}

\bibitem{2009-AbanovBettelheimWiegmann}
    A. G. Abanov, E. Bettelheim and P. Wiegmann, J. Phys. A: Math. Theor. {\bf 42}, 135201 (2009).
  \textit{Integrable hydrodynamics of Calogero-Sutherland model: bidirectional Benjamin-Ono equation.}



\bibitem{1995-Polychronakos}
    A. P. Polychronakos, Phys. Rev. Lett. \textbf{74}, 5153 (1995).
     \textit{Waves and Solitons in the Continuum Limit of the Calogero-Sutherland Model.}


\bibitem{kul1}    
     A. G Abanov, A. Gromov and M. Kulkarni, Journal of Physics A: Mathematical and Theoretical, Volume 44, Number 29, 17 June 2011. 
 \textit{Soliton solutions of a Calogero model in a harmonic potential}

\bibitem{kul2}    
     F. Franchini, A. Gromov, M. Kulkarni and A. Trombettoni, J. Phys. A: Math. Theor. 48 (2015) 28FT01
 \textit{Universal dynamics of a soliton after a quantum quench}



\bibitem{calfunc}  M. Bruschi and F. Calogero, , SIAM J. Math. Anal. 21(1990), 1019-1030
 \textit{General analytic solution of certain functional equations of addition type}




\bibitem{Szego-1975}
For a review and original references see G. Szeg\"o, \textit{Orthogonal Polynomials.}, fourth ed., American Mathematical Society, Providence, RI, 1975.

\bibitem{forrester}
    P. J. Forrester and J. B. Rogers, SlAM J. MATH. ANAL. Vol. 17, No. 2, March 1986
  {\it Electorstatics and the zeros of the classical polynomials.}

\bibitem{orive}
	R. Orive and Z. Garc�a, J. Comp. Appl. Math. \textbf{235}, 1065-1076 (2010).
  \textit{On a class of equilibrium problems in the real axis.}



\bibitem{mehta}
   M.L.Mehta, Random matrices, Third edition. Pure and Applied Mathematics (Amsterdam), 142. Elsevier/Academic Press, Amsterdam, 2004.

\bibitem{Stone}
Michael Stone, Inaki Anduaga and Lei Xing
Journal of Physics A: Mathematical and Theoretical, Volume 41, Number 27. \textit{The classical hydrodynamics of the Calogero–Sutherland model}

%%%%%% end of bib %%%%%

%\bibitem{i1}
%F. Calogero and C. Marchioro
%Phys. Rev. Lett. 27, 86 – Published 12 July 1971
%\\textit{Exact Ground State of Some One-Dimensional N-Body Systems with Inverse ("Coulomb-Like") and Inverse-Square ("Centrifugal") Pair Potentials}
%
%\bibitem{i2}
%F. Calogero, SIAM J. Math. Anal. 21(1990), 1019-1030
%Exactly solvable one dimensional many-body problems, Lett. Nuovo Cimento 13 (1975), 411-416
%\\textit{M. Bruschi and F. Calogero: General analytic solution of certain functional equations of addition type,  functional equation and solutions}
%
%
%\bibitem{AndricBardekJonke-1995}
%    I. Andri\'c, V. Bardek, L. Jonke, Phys. Lett. \textbf{B 357},
%   374 (1995).
%   \\ {\it Solitons in the Calogero-Sutherland collective-field model.}
%
%\bibitem{nikita}
%   V. Fock, A. Gorsky, N. Nekrasov, V. Rubtsov, JHEP 0007 028 (2000)
% \\ {\it Duality in Integrable Systems and Gauge Theories.}
%

%
%\bibitem{LECHTENFELD}
%    O. Lechtenfeld, Int. J. Mod. Phys. A \textbf{7}, 7097-7118 (1992).
% \\ {\it Semiclassical Approach to Finite-N Matrix Models.}
%
%
%\bibitem{itoi}
%    C. Itoi, Nucl. Phys. \textbf{B 493}, 651-659 (1997).
% \\ {\it Universal wide orthogonal, unitary correlators in non-gaussian and symplectic random matrix ensembles.}
%
%\bibitem{1980-Krichever}
%	I. M. Krichever, Funct. Anal. Appl. \textbf{14}, 282-290 (1980).
% \\ \textit{Elliptic solutions of the Kadomtsev-Petviashvhili equation and integrable systems of particles.}
%
%
%\bibitem{Ruijsenaars-1994}
%	S. Ruijsenaars,  2, Publ. RIMS Kyoto Univ. \textbf{30}, 865  (1994);
% \textit{Action-angle maps and scattering theory for some finite-dimensional integrable systems};
%	3, Publ. RIMS Kyoto Univ. \textbf{31}, 247  (1995);
%\textit{Action-angle maps and scattering theory for some finite-dimensional integrable systems.}
%
%
%\bibitem{1978-KKS}
%	D. Kazhdan, B. Kostant, and S. Sternberg, Commun. Pure Appl. Math. \textbf{31}, 481-507 (1978).
% \\ \textit{Hamiltonian Group Actions and Dynamical Systems of Calogero Type.}
%
%
%\bibitem{wilson}
%   G. Wilson, Invent. math. 133, 1-41 (1998)
% \\ {\it Collisions of Calogero-Moser particles and an adelic Grassmannian.}
%
%\bibitem{kasman}
%   A. Kasman, Commun. Math. Phys. 172, 427-448 (1995)
% \\ {\it Bispectral KP Solutions and Linearization of Calogero-Moser Particle Systems.}
%
%\bibitem{dui}
%   J. J. Duistermaat and F. A. Grunbaum, Commun. Math. Phys. 103, 177-240 (1986)
%   \\ {\it Differential Equations in the Spectral Parameter.}
%
%
%\bibitem{LPLM}
%	M. Bruschi and F. Calogero, Lett.  Nuovo Cimento \textbf{24}, 601-604 (1979).
% \\ \textit{Eigenvectors of a Matrix Related to the Zeros of Hermite Polynomials.}

%%%%%%%%%%%%%%%%%%%%%%%%%%%%%
\end{thebibliography}
\end{document}